\documentclass[%
	reprint,
 amsmath,amssymb,
 aps,
floatfix,
]{revtex4-1}

\usepackage{graphicx}
\usepackage{dcolumn}
\usepackage{bm}
\usepackage{amsbsy}
\usepackage{amsmath}
\usepackage{cancel}

\begin{document}


\title{Effect of compressibility and aspect ratio on performance of long elastic seals}

\author{B. Druecke}
	\email{bdruecke@mit.edu}
\author{D. M. Parks}
\author{A. E. Hosoi}%
\affiliation{%
 Massachusetts Institute of Technology
}%

\date{\today}

\begin{abstract}
Recent experiments show no statistical impact of seal length on the performance of long elastomeric seals in relatively smooth test fixtures. Motivated by these results, we analytically and computationally investigate the combined effects of seal length and compressibility on the maximum differential pressure a seal can support. We present a Saint-Venant type analytic shear lag solution for slightly compressible seals with large aspect ratios, which compares well with nonlinear finite element simulations in regions far from the ends of the seal. However, at the high- and low-pressure ends, where fracture is observed experimentally, the analytic solution is in poor agreement with detailed finite element calculations. Nevertheless, we show that the analytic solution provides far-field stress measures that correlate, over a range of aspect ratios and bulk moduli, the calculated energy release rates for the growth of small cracks at the two ends of the seal. Thus a single finite element simulation coupled with the analytic solution can be used to determine tendencies for fracture at the two ends of the seal over a wide range of geometry and compressibility. Finally, using a hypothetical critical energy release rate, predictions for whether a crack on the high-pressure end will begin to grow before or after a crack on the low-pressure end begins to grow are made using the analytic solution and compared with finite element simulations for finite deformation, hyperelastic seals.

\end{abstract}

\pacs{Valid PACS appear here}
\maketitle

\section{Introduction}
\begin{figure}[h!t]
	\centerline{
		\includegraphics[trim={0.5cm 24.2cm 8.0cm 0.7cm},clip,scale=0.65]{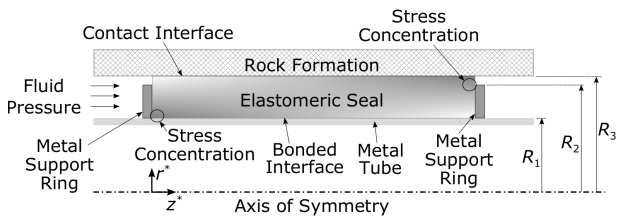}
	}
	\caption{Cross-sectional schematic of a typical hydraulic seal for oil and gas applications illustrating regions of high stress concentration at the high-pressure and low-pressure ends where material fracture is observed.}
	\label{Fig: Seal Schematic}
\end{figure}

There has been much work starting in the early part of this century on the use of gels as seals. Much of the interest was stimulated by the work of Beebe \cite{beebe2000functional}, who used them as pH-responsive valves in microfluidic devices. More recently, there has been a great deal of interest in their commercial use in oil and gas applications as seals for multi-stage hydraulic fracturing. A schematic of such a seal is shown in Fig. \ref{Fig: Seal Schematic}. A flurry of research related to this application has been conducted examining swelling kinetics \cite{cai2010force,chester2011thermo,lou2012swellable,liu2013kinetics,lou2014kinetics}, novel composite formulations \cite{robisson2013reactive,han2014swellable,han2014novel,qu2016inhomogeneous}, and fundamental mechanisms of sealing \cite{liu2014elastic,druecke2015large,wang2017extrusion}. However, nearly all of these studies assume the solvent and matrix phases of the gels to be incompressible, thereby rendering the sealing element incompressible aside from volumetric changes occurring via net imbibition of solvent from ambient.

It is undoubtedly true that the bulk modulus is much greater than the shear modulus for the polymer matrix in these gels, which is the reason for treating them as incompressible. For a conventional polybutadiene rubber, the bulk modulus is on the order of 2 GPa and the small-strain Young's modulus is on the order of 100 kPa \cite{tabor1994bulk}. Likewise the solvent phase imbibed by the matrix, consisting either of water or short-chain hydrocarbons, has a bulk modulus also on the order of gigapascals and, as a Newtonian fluid, has zero shear modulus. The fact that the bulk modulus is much larger than the shear modulus means that volume-preserving (isochoric) deformations often dominate dilational deformations, and treating these materials as perfectly incompressible is usually reasonable.

However, seal engineers are already aware that bulk modulus is important in accurately understanding the performance of seals in highly confined geometries. The reason is that, in these confined spaces, the isochoric deformations are often small, meaning the dilational deformations are of similar importance to volume-preserving deformations. While this is important for seals with aspect ratios near unity such as O-rings, it is exacerbated as the aspect ratio becomes long such that the dimension parallel to the axis of symmetry is much larger than the radial dimension perpendicular to the the axis. This is exactly the form factor used for swell packers for oil and gas applications.

The conventional wisdom in the oil and gas industry is that the critical differential pressure (the fluid pressure required for leakage) scales linearly with the length of the seal \cite{lemme2012swell}. However, recent laboratory experiments in relatively smooth-walled pressure vessels show contrary results. Motivated by these experimental observations, we investigate the interplay between seal length and compressibility and show how these variables affect performance of the seal.

We first briefly summarize the results of the motivating experiments. We then present an approximate, Saint-Venant type, analytic solution for the stress, strain and displacement fields in a long, slightly compressible seal for both axisymmetric and plane strain geometries.  Comparison with finite element simulations for large deformation, hyperelastic materials reveals that the linear, Saint-Venant solution is reasonable in the bulk of the seal away from the high-pressure and low-pressure ends. Despite the fact that the approximate solution is inaccurate near the two ends where failure is observed in experiments, we show that stress measures from the approximate solution are correlated with the energy release rates for the growth of cracks in these regions. These correlations, obtained by a single finite element simulation, can then be used to predict how seal length and bulk modulus affect the tendency to fracture at the two ends of the seal. The results show that increasing the bulk modulus or making the seal shorter promotes failure at the low-pressure end, and conversely, making the seal longer or more compressible promotes seal failure on the high-pressure end. Finally, for a particular choice of critical energy release rates, a phase plot is created showing when the seal will fail at the high-pressure end and when it will fail at the low-pressure end as a function of seal aspect ratio and ratio of elastic moduli. For the case in which friction is assumed to be small, increasing the length of a seal beyond a given scale does not lower the energy release rate, prevent the growth of cracks, nor improve the performance of seals.

\section{Motivation}
\begin{figure}[h!t]
	\centerline{
		\includegraphics[trim={0 13cm 10cm 1.0cm},clip,scale=0.65]{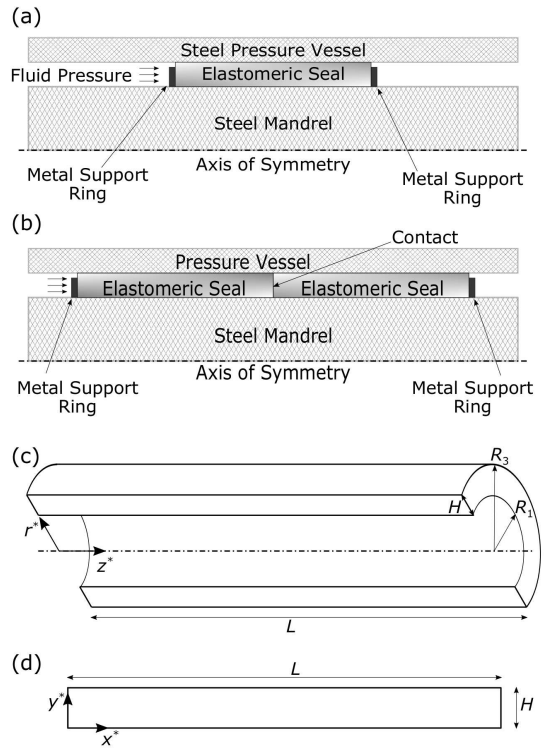}
	}
	\caption{Cross-sectional schematic of experimental configurations to investigate the effect of length on critical differential pressure. (a) Control configuration comprised of a single sealing element bonded on inner surface. (b) Two sealing elements stacked adjacent to each other. Metal rings constrain the ends of the assembly. The interface between the two sealing elements is a contact surface. (c) Geometry of the axisymmetric domain illustrating key dimensions. (d) Geometry of the plane strain domain.}
	\label{Fig: Stacking Schematic3}
\end{figure}

This work is motivated by a set of experiments examining the effect of seal length on the maximum differential pressure that the seal can support. Experiments were carried out with a 1/4 scale model of a commercial oilfield packer. A single seal element, shown schematically in Fig. \ref{Fig: Stacking Schematic3}, of length $L = 10.2 \mathrm{\ cm}$, inner diameter $D_{1} = 3.49 \mathrm{\ cm}$, support ring diameter $D_{2} = 4.61 \mathrm{\ cm}$ and swollen diameter $D_{3} = 4.83 \mathrm{\ cm}$ was able to support a maximum differential pressure of $\Delta P_{\mathrm{crit}} = 9.65 \pm 0.42$ MPa. The results of this control configuration were compared with those for two seal elements, each of length $L = 10.2 \mathrm{\ cm}$, stacked adjacent to each other with adjacent ends of the sealing elements in contact (but unbonded), as depicted in Fig. \ref{Fig: Stacking Schematic3}.b. This stacked configuration with twice the length of elastomer was able to support a maximum differential pressure of $\Delta P_{\mathrm{crit}} = 9.53 \pm 0.66$ MPa. Despite the fact that the seal was twice as long, the maximum differential pressure was not significantly different from the control configuration. Additional details of the experiments are found in Supplementary Material \S 1 at [URL will be inserted by publisher].

\section{Analysis}
	\subsection{Model Derivation}
Here we present the derivation of the approximate solution in plane strain geometry; the derivation for axisymmetric geometry is found in Appendix \ref{Appendix: Cylindrical Coordinates}. For small displacements $\mathbf{u}^{*} = u^{*} \left( x^{*}, y^{*} \right) \hat{\mathbf{e}}_{1} + v^{*} \left( x^{*}, y^{*} \right) \hat{\mathbf{e}}_{2} $ on the domain $\Omega = \left\{ \left( x^{*}, y^{*} \right) \in \mathbb{R}^{2} | 0 \le x^{*} \le L, 0 \le y^{*} \le H\right\}$ as depicted in Fig. \ref{Fig: Stacking Schematic3}.d, the governing equations for equilibrium linear elasticity are
\begin{equation}
	\boldsymbol{\nabla}^{*} \cdot \boldsymbol{\sigma}^{*} = \mathbf{0}
\end{equation}
where $\boldsymbol{\sigma^{*}} = 2 \mu \boldsymbol{\epsilon}^{*} + \lambda \mathrm{tr}\left( \boldsymbol{\epsilon^{*}} \right) \mathbf{I}$ is the Cauchy stress tensor, $\boldsymbol{\epsilon}^{*} = \frac{1}{2} \left( \boldsymbol{\nabla}^{*} \mathbf{u}^{*} + \left( \boldsymbol{\nabla}^{*} \mathbf{u}^{*} \right)^{T} \right)$ is the infinitesimal strain tensor, $\mu$ and $\lambda$ are Lam\'{e} moduli, $\mathbf{I}$ is the second-order identity tensor, and $\boldsymbol{\nabla}^{*}$ is the gradient operator with respect to the dimensional coordinates $\left(x^{*}, y^{*} \right)$. 

In typical applications, the seal is bonded on the inner surface to a rigid metallic substrate, thereby imposing a displacement-free boundary condition on this surface: $\mathbf{u}^{*} \left( x^{*}, y^{*} = 0 \right) = \boldsymbol{0}$. On the outer sealing surface, we impose a boundary condition in which the tangential traction and normal displacement are both zero, giving $\sigma^{*}_{12} = v^{*} = 0$ on $y^{*} = H$. Note that this is an idealization that is reasonable when there is little friction between the seal and the bounding surface, which is often true for cased holes and laboratory experiments. In the other limit, when the friction is very large, we can impose a zero tangential displacement boundary condition on the top surface at $y^{*} = 2 H$, giving rise to a shear-free symmetry plane at $y^{*} = H$, and the current formulation is still valid for plane strain. On the high-pressure end, fluid pressure $p_{0}$ is applied giving a compressive normal traction, $\sigma^{*}_{11} = -p_{0}$, and zero tangential traction, $\sigma^{*}_{12} = 0$, on $x^{*} = 0$. Finally, the physical boundary condition on the low-pressure end is complicated because there is contact between the seal and a rigid metal support ring that partially fills the annular gap between the radius of the inner tube, $R_{1}$, and the radius of the hole, $R_{3}$, in Fig. \ref{Fig: Seal Schematic}. Over the extrusion gap between the support ring ($R_{2}$) and outer hole ($R_{3}$), there is no traction on the low-pressure end of the seal. In order to make the problem analytically tractable, we parameterize this low-pressure boundary condition as an effective linear spring with stiffness $k^{*}$ such that $\sigma_{11}^{*} + k^{*} u^{*} = 0$ and $\sigma_{12}^{*} = 0$ on $x^{*} = L$. The effective stiffness is a function of the geometry of the low-pressure support, having values that range from $k^{*} = 0$ if there is no support to $k^{*} \rightarrow \infty$ if the metal support ring entirely fills the annular gap and the seal is perfectly supported. In practice, this stiffness can be determined by a single finite element simulation for a given geometry and shear modulus. Details of this calculation are given in Supplementary Material \S 2 at [URL will be inserted by publisher].

The entire premise of this work is that bulk deformation is as important or more important than isochoric deformation. We therefore scale the displacements with their bulk deformations
\begin{equation}
	u^{*} = \frac{p_{0} L}{K} u
	, \ 
	v^{*} = \frac{p_{0} H}{K} v
	, \
	x^{*} = L x
	, \
	y^{*} = H y
\end{equation}
where $K = \left( 2 \mu + 3 \lambda \right)/3$ is the bulk modulus. The motivation for this scaling comes from the relationship between the trace of the stress and strain tensors. The pressure in the solid, $p^{*} \left( x^{*}, y^{*} \right)$, is the spherical part of the stress tensor: $p^{*} = -\mathrm{tr} \boldsymbol{\sigma}^{*}/3 = -K \mathrm{tr} \boldsymbol{\epsilon}^{*} = -K \boldsymbol{\nabla}^{*} \cdot \mathbf{u}^{*} \sim p_{0}$. Then the non-dimensional equations to leading order are
	
\begin{subequations}
	\begin{equation}
		\frac{\partial}{\partial x} \left( \frac{\partial u}{\partial x} + \frac{\partial v}{\partial y} \right)
		+ \frac{\mu}{\lambda} \frac{L^{2}}{H^{2}} \frac{\partial^{2} u}{\partial y^{2}} = 0
	\end{equation}
	\begin{equation}
		\frac{\partial}{\partial y} \left( \frac{\partial u}{\partial x} + \frac{\partial v}{\partial y} \right) = -\frac{\partial p}{\partial y} = 0
		\ \ \Rightarrow \ p = p \left( x \right)
		\label{Eqn: Transverse Equilibrium}
\end{equation}
	\end{subequations}
	subject to the boundary conditions
	\begin{subequations}
	\begin{align}
	\left. \begin{array}{l}
	\sigma_{11} = -1 \\
	\sigma_{12} = 0
	\end{array}\right\} & \text{ on } x = 0
	\label{Eqn: Nondimensional Boundary Condition - HP}
\\
	\left. \begin{array}{r}
	\sigma_{11} + k u = 0 \\
	\sigma_{12} = 0
	\end{array}\right\} & \text{ on } x = 1
	\label{Eqn: Nondimensional Boundary Condition - LP}
\\
	\left. \begin{array}{c}
	u = 0 \\
	v = 0
	\end{array}\right\} & \text{ on } y = 0
	\label{Eqn: Nondimensional Boundary Condition - Inner}
\\
	\left. \begin{array}{r} 
	\displaystyle
	\sigma_{12} 
	= 0 
	\\ \displaystyle
	v = 0
	\end{array}\right\} & \text{ on } y = 1
	\label{Eqn: Nondimensional Boundary Condition - Outer}
	\end{align}
	\end{subequations}
	where each stress tensor component is normalized with the applied pressure $\boldsymbol{\sigma} = \boldsymbol{\sigma}^{*}/p_{0}$ and where $k \equiv k^{*} L/K$ is the nondimensionalized stiffness of the low-pressure support ring. We have assumed slight compressibility, $\mu / \lambda \ll 1$, and large aspect ratio, $L / H \gg 1$, and the corresponding small terms have been neglected.\\
	
	These equations are analogous to steady \emph{compressible} Poiseuille flow in a rectangular channel of height $2H$. Eqn. (\ref{Eqn: Transverse Equilibrium}) requires the pressure to be a function of axial position alone: $p = p \left( x \right)$. The pressure drop along the length of the seal is due to the no-slip constraint on $y = 0$ and the corresponding shear stress at this boundary. In the Poiseuille analogy, the low-pressure boundary condition can be thought of as a restrictive orifice, through which the fluid must flow, thereby providing a pressure drop proportional to flow rate. 
	
	Upon integration of the longitudinal equilibrium equation, application of boundary conditions at $y = 0$ and $y = 1$, substitution back into $p = -\boldsymbol{\nabla} \cdot \mathbf{u}$, integration, and application of the shear-free boundary condition on $y=1$, we obtain
	\begin{subequations}
		\begin{align}
		u \left( x , y \right) = & \frac{3}{\alpha^{2}} \left( \frac{1}{2} y^{2} - y \right) \frac{dp}{dx}
\\
		v \left( x, y \right) = &\left( -\frac{1}{2} y^{3} + \frac{3}{2} y^{2} - y \right) p \left( x \right)
\\
		p \left( x \right) = &C_{1} \cosh \alpha x + C_{2} \sinh \alpha x
		\end{align}
		\label{Eqn: Outer Solution}
	\end{subequations}
	where
	\begin{equation}
		\alpha \equiv \sqrt{3 \frac{\mu}{\lambda}} \frac{L}{H} = \sqrt{\frac{3 \left( 1 - 2 \nu \right)}{2 \nu}} \frac{L}{H}
	\end{equation}
	and $C_{1}$ and $C_{2}$ are constants of integration. Note that the parameter $\alpha$ is the key scaling parameter indicating the effect of system compressibility on the performance of the seal. As either $\mu/\lambda$ or $L^{2}/H^{2}$ become larger, the system becomes effectively more compressible, and a larger fraction of the fluid pressure applied to the seal is transmitted via shear to the basepipe, and a correspondingly smaller amount is transmitted to the low-pressure end of the seal.
	
	As in classic boundary layer problems \cite{bender1999advanced} where the outer solution is unable to satisfy all the requisite boundary conditions, the outer solution in Eqn. (\ref{Eqn: Outer Solution}) identically satisfies the boundary conditions on $y^{*} = 0$ and $y^{*}=H$ given by Eqn. (\ref{Eqn: Nondimensional Boundary Condition - Inner}) and (\ref{Eqn: Nondimensional Boundary Condition - Outer}), but has only two free parameters, $C_{1}$ and $C_{2}$, with which to attempt to satisfy four total boundary conditions on the surfaces $x^{*} = 0$ and $x^{*}= L$ required by Eqn. (\ref{Eqn: Nondimensional Boundary Condition - HP}) and (\ref{Eqn: Nondimensional Boundary Condition - LP}). Because we do not have analytic inner solutions in the boundary layer regions $x^{*} \lesssim H$ and $x^{*} \gtrsim L-H$ with which to match, we choose to satisfy the normal stress boundary conditions in a weak, Saint-Venant sense \cite{barber2010elasticity} on either end. The axial displacement and axial normal stress, averaged over the cross-sectional area, are
	\begin{scriptsize}
	\begin{subequations}
		\begin{equation}
			\overline{u} \left( x \right) = \int_{0}^{1}{u dy} 
			= -\frac{1}{\alpha} \left( C_{1} \sinh \alpha x + C_{2} \cosh \alpha x \right)
		\end{equation}
\vspace{-5mm}
		\begin{equation}
			\overline{\sigma}_{11} \left( x \right) = \int_{0}^{1}{\sigma_{11} dy}
			= -\frac{\lambda + 2 \mu}{K} \left( C_{1} \cosh \alpha x + C_{2} \sinh \alpha x \right).
		\end{equation}
	\end{subequations}
	\end{scriptsize}\noindent
	(Note that in the axisymmetric case in Appendix \ref{Appendix: Cylindrical Coordinates}, the integrals are evaluated using the differential cross-sectional area, $dA^{*} = 2 \pi r^{*} dr^{*}$.) Then applying the normal boundary conditions $\overline{\sigma}_{11} = -1$ on $x = 0$ and $\overline{\sigma}_{11} + k \overline{u} = 0$ on $x = 1$ gives
	\begin{equation}
		C_{1} = \frac{K}{\lambda + 2 \mu}, \ \ 
		C_{2} = -\frac{\cosh \alpha + k \frac{K}{\alpha \left( \lambda + 2 \mu \right)} \sinh \alpha}
		{\frac{\lambda + 2 \mu}{K} \sinh \alpha + k \frac{1}{\alpha} \cosh \alpha}
		\label{Eqn: Coefficients}
	\end{equation}
	where $k$ was previously defined to be the nondimensional stiffness of the low-pressure support. The resulting stress tensor components are
\begin{subequations}
	\begin{align}
			\sigma_{11}  = 
			&
			p \left( x \right) 
			\left[ 
				- \frac{\lambda}{K} 
				+ \frac{\mu}{K} \left( 3 y^{2} - 6 y \right)
			\right]
		\\
			\sigma_{22}  = 
			&
			p \left( x \right)
			\left[
				- \frac{\lambda}{K} 
				+ \frac{\mu}{K} \left( -3 y^{2} + 6 y - 2 \right) 
			\right]
		\\
			\sigma_{33}  =
			&
			- \frac{\lambda}{K} p \left( x \right)
		\\
			\sigma_{12}  = 
			&
			\frac{H}{L} \frac{dp}{dx}
			\left[ \frac{\lambda}{K} \left( y - 1 \right) + \frac{\mu}{K} \left( - \frac{1}{2} y^{3} + \frac{3}{2} y^{2} - y \right) \right].
		\end{align}\noindent
		\label{Eqn: Approximate Analytic Solution Summary}
	\end{subequations}

\noindent
	The approximate solution given in Eqn. (\ref{Eqn: Outer Solution}), (\ref{Eqn: Coefficients}) and (\ref{Eqn: Approximate Analytic Solution Summary}) is the outer asymptotic solution valid to leading order in the small parameter $\epsilon \equiv \mu/\lambda$. By the definition of bulk modulus, $K \equiv \lambda + 2 \mu / 3$, the ratio $\lambda / K = \mathcal{O} \left( \epsilon^{0} \right)$ is leading order. Some higher-order terms containing $\mu / K = \mathcal{O} \left( \epsilon^{1} \right)$ have been retained in Eqn. (\ref{Eqn: Approximate Analytic Solution Summary}) because they arise naturally from term proportional to the shear modulus, $\mu$, in $\boldsymbol{\sigma}^{*} = 2 \mu \boldsymbol{\epsilon}^{*} + \lambda \mathrm{tr} \left( \boldsymbol{\epsilon}^{*} \right) \mathbf{I}$, despite the solution for the displacements being valid only to leading order.
	
	\subsection{Validation}
\begin{figure}[h!t]
          \centerline{
              \includegraphics[trim={0cm 0cm 1cm 0.0cm},clip=true,width=3.0in]{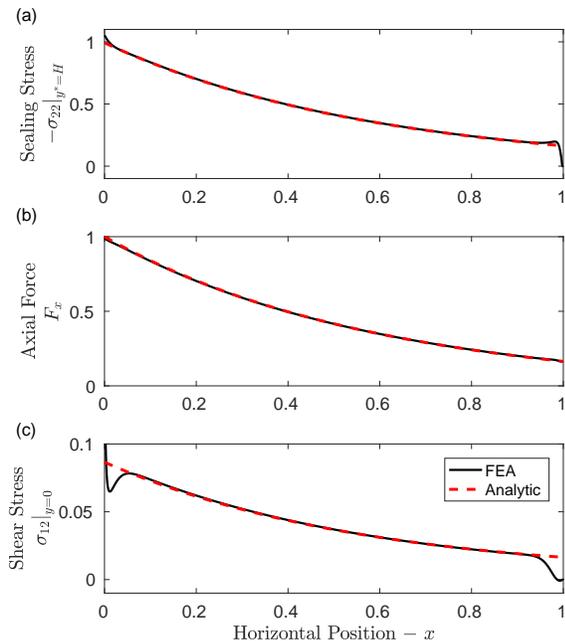}
           }
		\caption{Comparison between finite element results and approximate analytic solution of several variables relevant for seal performance. (a) Sealing stress between upper surface of seal and top rigid surface at $y^{*} = H$. (b) Axial force, $F_{x} = \left( p_{0} H \right)^{-1} F^{*}_{x}$, along the length of the seal. (c) Shear stress at interface where seal is bonded to inner rigid surface at $y^{*} = 0$.}
		\label{Fig: Stress vs Axial Position}
	\end{figure}

	The asymptotic solution does not satisfy the shear-free boundary conditions at $x^{*} = 0$ and $x^{*} = L$, and is therefore valid only in the bulk of the domain, $H \lesssim x^{*} \lesssim L - H$, sufficiently far from the boundary layers at the ends, which are of thickness $\mathcal{O} \left( H \right)$. The utility of the analytic solution can be seen by comparing key dependent variables with those predicted by more realistic finite element solutions.
	Fig. \ref{Fig: Stress vs Axial Position} compares results of the approximate analytic solution to detailed finite element calculations (finite deformation, neo-Hookean hyperelastic material) for a case when the aspect ratio is $L/H = 20$, the ratio of small strain elastic moduli are $\mu / \lambda = 2.5 \times 10^{-3}$ and the low-pressure support ring fills three quarters of the gap between the inner and outer surfaces of the seal. Note that when referring to material properties in finite element simulations, we use $\mu$ and $\lambda$ to denote the small-strain limits of the elastic moduli for finite element simulations of finite deformation hyperelasticity. 

	Three quantities of interest to seal engineers are plotted. The first of these is the sealing stress between the outer surface of the seal and the rigid surface against which it seals. Lorenz and Persson \cite{lorenz2009leak} have shown that leak rate past the seal increases with a decrease in this contact stress, and the normal contact traction is of primary importance in modeling the flow of fluid between deformable surfaces, as done using the pressure penetration boundary condition in Abaqus/Standard \cite{Abaqus}. Sealing stress is well-predicted by the analytic solution except in the boundary layer regions near the ends given by $x \lesssim H/L$ and $x \gtrsim 1 - H/L$. Note that the present solution does not include the effect of pre-compression of the seal, which is required for sealing, but, for a linear system, pre-compression can be accounted for by superposition.
	
Similarly, as one might intuit and as we show below, the transfer of axial force,
\begin{equation}
	\displaystyle F_{x}^{*} \left( x^{*} \right) \equiv -\int_{0}^{H}{\sigma_{11}^{*} \left( x^{*}, y^{*} \right) dy^{*}},
\end{equation}
along the length of the seal from the high-pressure end to the low-pressure end is important for predicting the onset of failure on both ends. Note that the force is defined to be positive when acting in the $+ \hat{\mathbf{e}}_{1}$ direction on a cut through the body with outward unit normal $\hat{\mathbf{n}} = -\hat{\mathbf{e}}_{1}$. (This is simply for convenience, because the axial stress, $\sigma_{11}^{*}$, is compressive and therefore negative.) As shown in Fig. \ref{Fig: Stress vs Axial Position}.b, the analytic solution is able to predict axial force along the length of the seal.

Finally, one might be interested in shear stress along the inner surface where the seal is bonded to a rigid substrate. By equilibrium in the longitudinal direction and the shear-free boundary condition on the upper surface, the shear traction is equal to the slope of the axial force curve.
\begin{equation}
	\int_{0}^{H}{\left( \frac{\partial \sigma_{11}^{*}}{\partial x^{*}} + \frac{\partial \sigma_{12}^{*}}{\partial y^{*}} \right) dy^{*} = 0}
	\ \Rightarrow \
	\frac{d F_{x}^{*}}{dx^{*}} = \left. \sigma_{12}^{*} \right|_{y^{*}=0}.
	\label{Eqn: Axial Stress Equation}
\end{equation}
Fig. \ref{Fig: Stress vs Axial Position}.c shows that the agreement is reasonable, except in the boundary layer regions at the ends.

	\subsection{Failure Initiation}
	In experimental observations of typical oilfield seals, large cracks are often observed on the low-pressure end of the seal. Additionally, delamination between the seal and the rigid substrate is sometimes observed on the high-pressure end \cite{druecke2015large}, as depicted schematically in Fig. \ref{Fig: Schematic - Matched Asymptotics}. In what follows, we presume that there are small preexisting cracks in the material, both on the high-pressure end at the interface between the seal and the substrate and at the low-pressure end near the extrusion gap. The precise geometry of these cracks is generally unknown. We investigate, through the use of finite element simulations, the combined effects of seal length and bulk modulus on the tendency for growth of these cracks. The parameter of interest for crack mechanics is the energy release rate, $\mathcal{G}$, which gives the incremental decrease in elastic strain energy per incremental increase in crack surface area as the crack advances. Classic energy arguments pioneered by Griffith state that the crack will advance when the energy release rate exceeds a critical value, $\mathcal{G}_{\mathrm{crit}}$, which is a material property, such that
	\begin{equation}
		\mathcal{G} \ge \mathcal{G}_{\mathrm{crit}}.
	\end{equation}
	To investigate these tendencies for crack propagation, we conduct a sequence of finite element simulations for the finite deformation of a neo-Hookean material. The bulk modulus and length of the seal are varied. We use the finite element program Abaqus/Standard to calculate the energy release rate \cite{parks1977virtual} for the growth of an interfacial crack of length 10 $\mu$m on the high-pressure end. We use the same simulations to calculate the energy release rate for the growth of each crack in an array of 51 cracks on the low-pressure end. Each crack in this array extends perpendicularly inward from the low-pressure surface, has a length of 10 $\mu$m, and a spacing from its nearest neighbors of $s = 20$ $\mu$m. Although this close spacing of cracks leads to shielding (see Supp. Info. \S 3), giving a lower energy release rate for a particular crack than if there were no neighboring cracks, the crack geometry and shielding effect are held fixed in all simulations, and the shielding does not change the trends observed in the dependence of energy release rates on length and bulk modulus. Details of the simulations are provided in the Supplementary Material at [URL will be inserted by publisher].
	
	\subsubsection{High-Pressure End}
	In physical seal systems, there is a radial compressive stress that is needed to prevent the flow of fluid betweeen the seal and its mating surfaces. This stress acts perpendicular to the plane of the small interfacial crack on the high-pressure end and serves to close the crack. Therefore, the driving force for crack propagation is the shearing between the seal and the rigid substrate, which gives rise to a Mode II shear failure at the interface. Although the behavior of interfacial cracks between dissimilar materials is nontrivial \cite{rice1988elastic}, we expect both the energy release rate and the Mode II stress intensity factor to be proportional to the shear stress at the high-pressure corner
		\begin{equation}
			\mathcal{G}_{\mathrm{HP}} \sim \frac{K_{\mathrm{II}}^{2}}{E'} \sim \frac{\sigma_{12}^{2} a}{E'}
			\label{Eqn: HP Energy Release Rate}
		\end{equation}
		where $a$ is the crack length, $\mathcal{G}_{\mathrm{HP}}$ is the energy release rate for the growth of a crack at the high-pressure end, $E' = E/(1-\nu^2) = 4 \mu \left( \lambda + \mu \right) / \left( \lambda + 2 \mu \right) \approx 4 \mu$ is an effective Young's modulus for plane strain, and $\sigma_{12}$ is some characteristic (far-field) shear stress evaluated near the high-pressure corner at $\left(x^{*}, y^{*} \right) = \left( 0, 0 \right)$. Despite the singular nature of the shear stress at the crack tip in linear elasticity theory, we take a coarse-grained approach and treat the shear stress ahead of the crack as the gradient of axial force given in Eqn. (\ref{Eqn: Axial Stress Equation}). Then we expect a quadratic dependence of energy release rate on gradient of force, as
\begin{equation}
	\mathcal{G}_{\mathrm{HP}} \sim \left( \left. \frac{dF_{x}^{*}}{dx^{*}} \right|_{x^{*} = 0} \right)^{2}.
\end{equation}
		
\begin{figure}[h!t]
  \centerline{\includegraphics[width=3.25in]{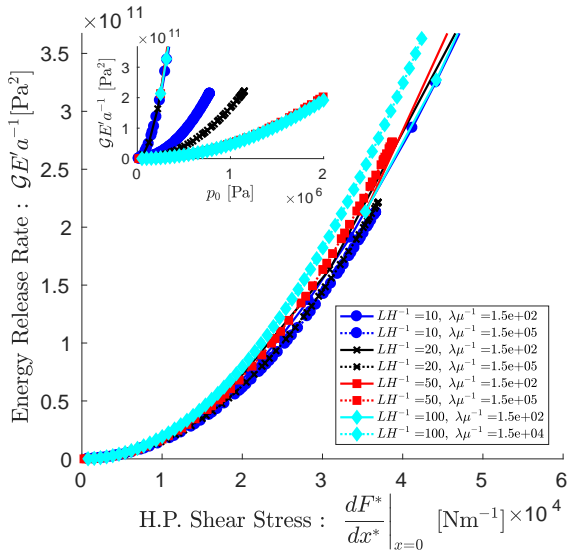}}
  \caption{The energy release rate for the growth of a crack at the high-pressure end depends on the average shear stress in this region, regardless of seal length or bulk modulus. The leading edge shear stress is given by the analytic solution as the gradient of axial force from Eq. (\ref{Eqn: Axial Stress Equation}). Inset: Energy release rate as function of pressure varies with geometry and material properties.}
	\label{Fig: HP Energy Release}
\end{figure}

Fig. \ref{Fig: HP Energy Release} shows the energy release rate computed by the finite element simulations versus the analytically predicted gradient of axial force at the leading edge, $\left. dF^{*}/dx^{*}\right|_{x = 0}$, for several aspect ratios and two drastically different bulk moduli. The results show that this scaling, and the approximate analytic solution for the axial force gradient correlate the computed energy release rates over a wide range of aspect ratios and ratios of elastic moduli.
	
Aside from the collapse of all the energy release rates versus shear stress, it is worthwhile to note that, in the inset of Fig. \ref{Fig: HP Energy Release}, all the energy release rate curves for highly compressible seals with the ratio of elastic moduli $\lambda/\mu = 150$ fall on top of each other, despite the widely varying aspect ratios. This indicates that the energy release rate for the growth of a crack on the high-pressure end of the seal is insensitive to seal length, when the bulk modulus is low and the seal is highly compressible. The reason is that, for highly compressible seals, the seal does not feel the effect of the low-pressure boundary condition, and the entirety of the applied differential pressure is transferred to the rigid substrate via shear. The seal is effectively infinitely long from the high-pressure crack's perspective. Thus, for a fixed bulk modulus, increasing seal length beyond a certain value does not affect the tendency for crack growth on the high-pressure end and does not improve seal performance.

	\subsubsection{Low-Pressure End}
\begin{figure}[h!t]
  \centerline{\includegraphics[width=1.0\columnwidth]{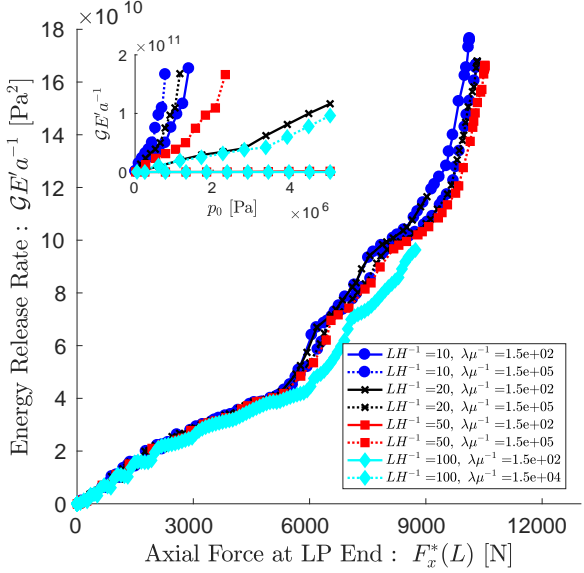}}
  \caption{Energy release rate for the growth of cracks on the low-pressure end. Inset gives energy release rate versus applied pressure on the high-pressure end. Primary graph shows that energy release rate is dependent only upon axial load transmitted to low-pressure end, regardless of length or material compressibility. The odd shape of the curves results from finite deformation (see Supplementary Material at [URL will be inserted by publisher]).}
  \label{Fig: LP Energy Release}
\end{figure}

	On the low-pressure end, we expect the energy release rate for the growth of a small crack to depend on the load transmitted from the high-pressure end along the length of the seal to the extrusion gap at the low-pressure end. The approximate analytic solution above provides a prediction of the distribution of axial load, $F^{*} \left( x^{*} \right)$, and we therefore attempt to scale energy release rate with transmitted axial force as
\begin{equation}
	\mathcal{G}_{\mathrm{LP}} \sim F_{x}^{*} (L)
\end{equation}
Fig. \ref{Fig: LP Energy Release} shows that, despite the rather unusual shape of the energy release rate versus pressure curves due to finite deformations, the energy release rate on the low-pressure end is well-predicted by the transmitted axial force, which is calculated from the approximate analytic solution.

The unusual shapes of both energy release rate versus applied pressure, $p_{0}$, and energy release rate versus transmitted axial force, $\left. F^{*}_{x} \right|_{x^{*} = L}$ are a manifestation of finite deformation in the simulations. As the pressure increases, the seal material is pushed into the extrusion gap. Therefore, a small preexisting crack originally located at the corner of the support ring where the stress concentration is highest is extruded into a region where the stress concentration is lower, and a new crack moves into the region of highest stress concentration. The curve of energy release rate is the locus of the instantaneously maximum (evaluated over all the cracks) energy release rate
\begin{equation}
	\mathcal{G}_{\mathrm{LP}} \left( p_{0} \right) = \max_{i}{\mathcal{G}_{\mathrm{LP},i} \left( p_{0} \right)}
\end{equation}
where $\mathcal{G}_{\mathrm{LP},i} \left( p_{0} \right)$ is the energy release rate of the $i^{\mathrm{th}}$ crack at applied pressure $p_{0}$. In typical linear elastic fracture mechanics, the energy release rate scales with the square of the applied load. Here, the scaling is initially weaker than that because finite deformation extrusion provides an additional degree of freedom, thereby relieving some of the increase in strain energy at the crack tips. 

Most importantly, we do not claim that the shape of this curve is univeral. Undoubtedly it depends on the geometry of the low-pressure support and the geometry of the cracks in the low-pressure region. However, the collapse of all of these curves over a range of aspect ratios and elastic moduli ratios, when scaled with transmitted axial load, is the key finding.

	\subsubsection{Qualitative Explanation of Observed Correlation}
	Here we provide a qualitative explanation for the reason the analytic solution provides good correlations of energy release rates despite having poor agreement with detailed finite element simulations at both ends of the seal. The reasoning is analogous to the theory of matched asymptotic expansions and conventional fracture mechanics arguments regarding ratios of length scales. We have already shown through finite element simulations that the deformations in the bulk of the seal are small, due to the highly confined nature of the seal, and have utilized Saint-Venant's principle to obtain a solution in the bulk without accurately resolving the details in the boundary layer regions at the ends of the seal. The bulk solution is insensitive to the detailed solution at the high-pressure end, and therefore is insensitive to the details of the stress and displacement fields surrounding the crack tip for a short interfacial crack. Similarly, on the low-pressure end, the boundary condition has been parameterized as a linear spring, and the bulk solution is therefore only sensitive to the overall stiffness of the low-pressure support and insensitive to the details of the crack tip stress and displacement fields. 

	On the other hand, the inner solutions (which we have only computed numerically) near the crack on the high-pressure end and near the cracks on the low-pressure end are highly dependent on the details of the crack geometry as well as the far-field loading of the crack. Although in this analysis we do not claim to know the detailed geometry of the cracks (crack length, spacing, etc.), it is clear that once this geometry is fixed, the only variables affecting the stress intensity factor and energy release rate are the far-field loading and material properties. The energy release rate for these cracks is dependent only on shear modulus or Young's modulus, and is not strongly affected by the bulk modulus, as illustrated by Eqn. (\ref{Eqn: HP Energy Release Rate}), where the Poisson ratio is always very near to its incompressible limit: $\nu \approx 1/2$. 

Having fixed the crack geometry and the shear modulus, the only remaining parameter affecting energy release rate is the far-field loading on the cracks, which is precisely what is provided by the approximate analytic solution in the bulk. Therefore, holding the shear modulus and local geometry around the cracks fixed, the outer analytic solution provides far-field loading on the crack and can be used to correlate energy release rates for the growth of cracks when the length and bulk modulus of the seal are varied.

\begin{figure*}[h!t]
			\centerline{\includegraphics[width=6.5in]{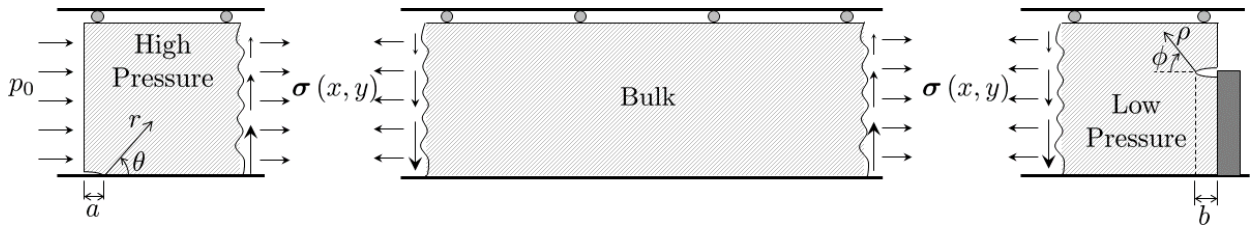}}
			\caption{Schematic illustrating the boundary value problem near the crack tips and the separation of the problem into boundary layers, on which the solution is evaluated numerically, and an outer region over which the approximate analytic solution is reasonable.}
			\label{Fig: Schematic - Matched Asymptotics}
		\end{figure*}

Fig. \ref{Fig: Schematic - Matched Asymptotics} schematically illustrates this separation of length scales and matching. On the high-pressure end, the far-field loading condition is dependent both on the applied pressure, $p_{0}$, and on the traction exerted by the bulk of the seal on the high-pressure end. For Mode II fracture, the far-field stress of importance is this gradient of axial force given by the difference between the applied pressure $p_{0}$ and the traction on the hypothetical cut dividing the high-pressure end from the bulk of the seal
\begin{displaymath}
	\sigma_{\infty,\mathrm{(HP)}} \sim \lim_{r \rightarrow \infty}{\frac{p_{0} + \int_{0}^{1}{\mathbf{t}_{\mathrm{HP}} \left( x, y \right) \cdot \hat{\mathbf{e}}_{x} dy}}{x}}
\end{displaymath}
where $r$ is the radial coordinate from the tip of the interface crack on the high-pressure end. However, from the theory of matched asymptotic expansions, we know that the limit of the outer, bulk solution as $x^{*} \rightarrow 0$ must equal the limit of the inner, local crack solution as $r \rightarrow \infty$.
%
	\begin{multline}
	\lim_{\frac{r}{a} \rightarrow \infty}{\boldsymbol{\sigma}_{\mathrm{HP}}} = \lim_{\frac{x^{*}}{L} \rightarrow 0}{\boldsymbol{\sigma}_{\mathrm{bulk}}}
\\
	\Rightarrow
	\sigma_{\infty,\mathrm{HP}} \sim \lim_{\frac{x^{*}}{L} \rightarrow 0}{\frac{-\sigma_{xx} \left( x^{*} = 0 \right) + \int_{0}^{1}{\sigma_{xx\mathrm{(bulk)}} \left( x^{*}, y\right) dy}}{x^{*}}}.
	\end{multline}
%
This is the gradient of average axial force, or, using Eqn. \ref{Eqn: Axial Stress Equation}, this is equivalent to the shear stress in the bulk predicted by the approximate analytic solution.

Similarly, on the low-pressure end, the far-field stress of importance is the traction in the horizontal direction applied far from the crack as $\frac{\rho}{b} \rightarrow \infty$ where $\rho$ is the radial coordinate from the tip of the low-pressure crack and $b$ is the length of the low-pressure crack. As in the analysis for the high-pressure end, the theory of matched asymptotic expansions states that the outer limit of stresses in the low-pressure solution must equal the inner limit of stresses in the bulk solution
\begin{displaymath}
	\lim_{\frac{x^{*}}{L} \rightarrow 1}{\boldsymbol{\sigma}_{\mathrm{(Bulk)}}} = \lim_{\frac{\rho}{b} \rightarrow \infty}{\boldsymbol{\sigma}_{\mathrm{(LP)}}}
\end{displaymath}
which indicates that the bulk solution evaluated at the low-pressure end gives the proper scaling of the far-field loading for a small crack at the low-pressure end.

\section{Discussion}
In the preceding section, we derived an approximate analytic solution for the distribution of stresses and displacements in a long, slightly compressible seal. We have also shown that the energy release rate for the growth of a small crack on the high-pressure end of the seal at the interface between the seal and the rigid substrate is dependent on the coarse-grained shear stress at this location. Finally, we have shown that the energy release rate for the growth of a small crack on the low-pressure end depends on the axial force transmitted from the high-pressure end to the low-pressure end by the seal. Here, we put together these pieces of information and use them to analyze the effects of bulk modulus and seal length on seal performance. Specifically, we show that decreasing the bulk modulus decreases the tendency for crack growth on the low-pressure end, which implies that the seal is more likely to begin to fail on the high-pressure end. Similarly, increasing the seal length makes the seal effectively more compressible and decreases the energy release rate for growth of a crack on the low-pressure end. We examine this behavior and compare analytic predictions with a small parametric finite element investigation illustrating these points. 

Note that the geometry of the low-pressure support for this finite element study differs from that used to generate Figs. \ref{Fig: HP Energy Release} and \ref{Fig: LP Energy Release}. Here, an extrusion gap of ten percent of the seal thickness is used, whereas the previous results were for an extrusion gap that was 25 percent of the seal thickness. The present results give a correspondingly stiffer low-pressure support than those of the preceding section. Further parameters for the finite element simulations are given in the Supplementary Material at [URL will be inserted by publisher].

The analytic solution above gave the distribution of axial and shear stresses in the body, which were used to correlate energy release rate in Figs. \ref{Fig: HP Energy Release} and \ref{Fig: LP Energy Release}. In dimensional form, the analytically predicted shear stress on the high-pressure end and the transmitted axial force on the low-pressure end were, respectively,
\begin{subequations}
	\begin{align}
\left. \frac{dF_{x}^{*}}{dx^{*}} \right|_{x^{*} = 0} = &
		-p_{0} \alpha \frac{H}{L} \frac{\lambda}{K} C_{2}
\\
F_{x}^{*} \left( x^{*} = L \right) = & 
		-p_{0} H \frac{3 \lambda + 6 \mu}{3 \lambda + 2 \mu} \left( C_{1} \cosh \alpha + C_{2} \sinh \alpha \right).
	\end{align}
	\label{Eqn: Stress Solutions}
\end{subequations}

\noindent
As stated above, given the material properties and geometry of the seal, this solution has a single free parameter, $k^{*}$, which characterizes the stiffness of the low-pressure support. For a neo-Hookean seal with a small strain shear modulus of $\mu = 6.13 \times 10^{5}$ Pa and geometry given in Table \ref{Table: Geometry}, the low-pressure support stiffness is found, via finite element analysis, to be $k^{*} = 5.59 \times 10^9 \pm 4.3 \times 10^7$ Pa/m. 

\begin{table}[h!t]
	\caption{Geometry for analysis of a seal with 10 percent extrusion gap}
	\label{Table: Geometry}
	\begin{tabular}{|l|c|}
		\hline
		Seal Thickness [m] & 0.025 \\ \hline
		Seal Length [m] & Variable \\ \hline
		Gauge Ring Thickness [m] & 0.0225 \\ \hline
		Gap [m] & 0.0025 \\ \hline
		Gauge Ring Corner Radius [m] & 0.000188 \\ \hline
		Crack Length [m] & $1.0 \times 10^{-5}$ \\ \hline
	\end{tabular}
\end{table}

\begin{figure}[h!t]
	\centerline{\includegraphics[scale=0.6]{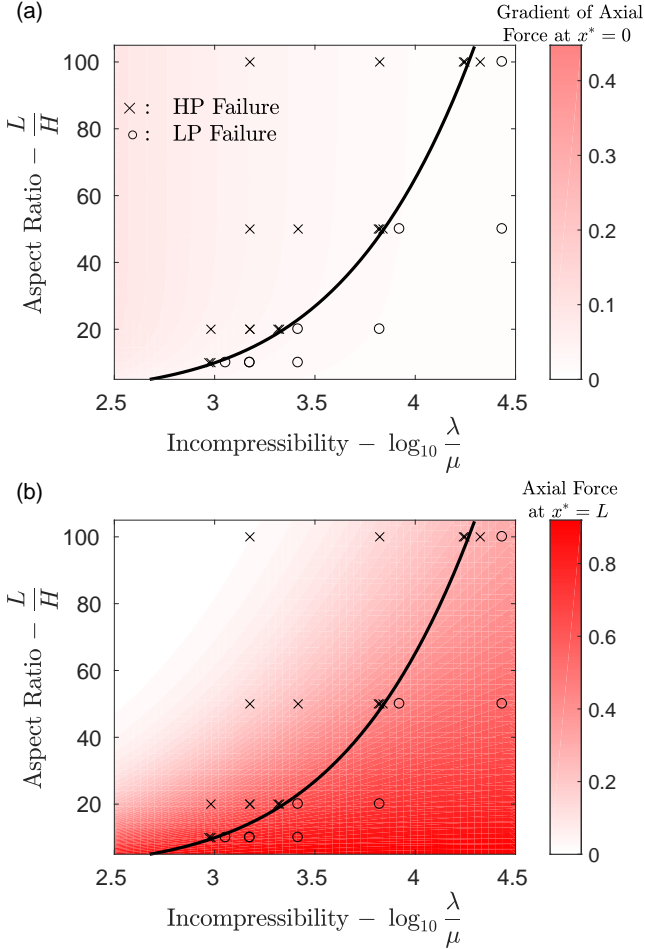}}
	\caption{Stress as a function of aspect ratio and ratio of elastic moduli. $\times$: Finite element simulations where failure first occurs on high-pressure (HP) end. $\circ$: Finite element simulations where failure first occurs on the low-pressure (LP) end. (a) Shear stress, $p_{0}^{-1} dF^{*}/dx^{*} \left( x=0\right)$, at the high-pressure end given by the approximate analytic solution. (b) Average axial stress, $p_{0}^{-1} H^{-1} F_{x}^{*} \left( x^{*} = L \right)$, at the low-pressure end. (Upper left region has no values because the parameter region gives unrealistically large values of $\alpha$ here, such that asymptotic solution is not valid.) Solid lines indicate contours along which the tendency for failure at high-pressure end relative to low-pressure end is constant for fixed value of $k$ and fixed critical energy release rates.}
	\label{Fig: Stress Contours}
\end{figure}

Using this value of $k^{*}$, the stresses in Eqn. (\ref{Eqn: Stress Solutions}) are known. Fig. \ref{Fig: Stress Contours} shows a plot of the normalized stress on the high-pressure end, $p_{0}^{-1} dF^{*}/dx^{*}$ and the normalized stress on the low-pressure end, $p_{0}^{-1} H^{-1} F_{x}^{*} \left( x^{*} = L \right)$, both as a function of bulk modulus and aspect ratio. The results show that, on the high-pressure end, the shear stress decreases with increasing bulk modulus because, as the bulk modulus increases, more of the applied load is supported by the low-pressure support and less by shear along the bonded interface. Similarly, as the length of the seal decreases, more load is transferred to the low-pressure support and less is supported by shear. The opposite trends in axial stress are observed at the low-pressure end of the seal.

In addition to the analytic solution, the preceding section demonstrated that the energy release rates on the high- and low-pressure ends depend on these stress measures. For the sake of clarity of discussion, let us denote the functional dependence of the energy release rates on these stress measures as
\begin{subequations}
	\begin{align}
	\mathcal{G}_{\mathrm{HP}} =& \phi_{\mathrm{HP}} \left( dF^{*}/dx^{*} \left( x^{*} = 0 \right); \mathrm{Crack \ Length}, \mu, \cdots \right)
\\
	\mathcal{G}_{\mathrm{LP}} =& \phi_{\mathrm{LP}} \left( F^{*} \left( x^{*} = L \right); \mathrm{Crack \ Length}, \mathrm{Gap}, \mathrm{\mu}, \cdots \right)
	\end{align}
	\label{Eqn: Energy Release Rate vs Stress}
\end{subequations}

\noindent
where the dependence on variables such as crack length could be further investigated using finite element analysis but are simply held fixed in the current study. The functional dependence of energy release rate on stress, as denoted by $\phi_{\mathrm{HP}}$ and $\phi_{\mathrm{LP}}$, can be determined from a single finite element analysis, as illustrated in Figs. \ref{Fig: HP Energy Release} and \ref{Fig: LP Energy Release} (see also Figs. SI-7 and SI-8), and then used for varying seal lengths and bulk moduli. 

Although this current investigation makes no attempt to experimentally characterize the critical energy release rate for growth of cracks on either end of a seal, using hypothetical values allows us to explore the effects of aspect ratio and bulk modulus. If we computationally determine the dependence of energy release rates on the stress measures of Eqn. (\ref{Eqn: Stress Solutions}), then we can graphically invert these functions to obtain stress as a function of energy release rate. Specifically, we are interested in stresses corresponding to critical energy release rates, $\mathcal{G}_{\mathrm{HP}_{\mathrm{crit}}}$ and $\mathcal{G}_{\mathrm{LP}_{\mathrm{crit}}}$ at which cracks begin to grow. Inverting Eqn. (\ref{Eqn: Energy Release Rate vs Stress}) while holding constant crack length, gap, etc., gives
\begin{subequations}
	\begin{align}
	\left. \frac{dF^{*}}{dx} \right|_{x^{*}=0,\mathrm{crit}} =& \phi_{\mathrm{HP}}^{-1} \left( \mathcal{G}_{\mathrm{HP}_{\mathrm{crit}}} \right)
\\
	\left. F^{*} \right|_{x^{*} = L, \mathrm{crit}} =& \phi_{\mathrm{LP}}^{-1} \left( \mathcal{G}_{\mathrm{LP}_{\mathrm{crit}}} \right).
	\end{align}
\end{subequations}
Furthermore, substituting into Eqn. \ref{Eqn: Stress Solutions} allows us to determine the pressure required to reach the critical energy release rate at each of the two ends of the seal.
\begin{subequations}
	\begin{align}
p_{\mathrm{HP}_{\mathrm{crit}}} =& \frac{\phi_{\mathrm{HP}}^{-1} \left( \mathcal{G}_{\mathrm{HP}_{\mathrm{crit}}} \right)}
						{-\alpha \frac{H}{L} \frac{\lambda}{K} C_{2}}
\\
p_{\mathrm{LP}_{\mathrm{crit}}} =& \frac{\phi_{\mathrm{LP}}^{-1} \left( \mathcal{G}_{\mathrm{LP}_{\mathrm{crit}}} \right)}
						{\frac{H \left( 2 \mu + \lambda \right)}{K} \left( C_{1} \cosh \alpha + C_{2} \sinh \alpha \right)}.
	\end{align}
\end{subequations}
The minimum of these two is the pressure at which the seal will begin to fracture. If the critical pressure on the high-pressure end is smaller, a crack on the high-pressure end will begin to grow before a crack on the low-pressure end will advance. Conversely, if the critical pressure on the low-pressure end is smaller, a crack there will begin to grow first. The point at which cracks on both the high-pressure end and low-pressure end are equally likely to advance is given by the equality of these two critical pressures, which can be solved for via a numeric root-finding algorithm, such as Newton-Raphson or \textsc{Matlab}'s \texttt{fminsearch}.

In the absence of knowledge of values of critical energy release rates, $\mathcal{G}_{\mathrm{HP}_{\mathrm{crit}}}$ and $\mathcal{G}_{\mathrm{LP}_{\mathrm{crit}}}$, we choose hypothetical values and examine the resulting behavior. Because of convergence issues and the inability to push the finite element simulations to high energy release rates for cracks with length of 10 $\mu$m, as given in Table \ref{Table: Geometry}, critical energy release rates of $\mathcal{G}_{\mathrm{HP}_{\mathrm{crit}}} = \mathcal{G}_{\mathrm{LP}_{\mathrm{crit}}} = 0.5 \text{ J}/\text{m}^{2}$ were chosen for the purpose of illustrating the combined effects of length and compressibility. Despite the fact that these are low for realistic systems, the qualitative behavior obtained from them is representative. Using plots analogous to those of Figs. \ref{Fig: HP Energy Release} and \ref{Fig: LP Energy Release} (which are contained in the Supplementary Material Figs. SM-7 and SM-8 at [URL will be inserted by publisher]) for the geometry of Table \ref{Table: Geometry}, the resulting critical shear stress values are $dF^{*}/dx^{*}_{x^{*} = 0,\mathrm{crit}} = \phi_{\mathrm{HP}}^{-1} \left( 0.5 \text{ J}/\text{m}^{2} \right) \approx 2.6 \times 10^{4}$ N/m and $\left. F^{*} \right|_{x^{*} = L, \mathrm{crit}} = \phi_{\mathrm{LP}}^{-1} \left( 0.5 \text{ J}/\text{m}^{2} \right) \approx 1.4 \times 10^{4}$ N. Then using these values and equating the critical pressure on the low-pressure end with that on the high-pressure end gives the solid black curve shown in the plots of Fig. \ref{Fig: Stress Contours}. This is the predicted curve on which failure at the low-pressure end is equally favorable as failure at the high-pressure end. To the left of this curve, stress on the high-pressure end increases, while stress on the low-pressure end decreases, making failure on the high-pressure end more favorable. Conversely, to the right of this curve, failure on the low-pressure end is more probable. 

Superposed on this plot are symbols denoting results of finite element simulations. In these finite element simulations, energy release rates were calculated for the growth of small cracks on both the high-pressure and low-pressure ends of the seal. The same failure criteria of $\mathcal{G}_{\mathrm{HP}_{\mathrm{crit}}} = \mathcal{G}_{\mathrm{LP}_{\mathrm{crit}}} = 0.5 \text{ J}/\text{m}^{2}$ were used for the finite element simulations, and once the computed energy release rate exceded the critical energy release rate, the seal was deemed to have failed, with the location of failure given by the end that first reached the critical energy release rate. In the plot, the symbols $\times$ denote simulations in which the failure first occurred on the high-pressure end and the symbols $\circ$ denote those simulations in which the failure first occurred on the low-pressure end. The predictions by the set of finite element simulations predicting whether the seal will fail at the high-pressure end or the low-pressure end agree well with the predictions made from the analytic solution.

These results are consistent with expectations regarding the effect of aspect ratio and compressibility. For a given aspect ratio, increasing the bulk modulus shifts the location of failure from the high-pressure end to the low-pressure end because a high bulk modulus means more axial load is transferred to the low-pressure end and less is transferred via shear to the rigid substrate. Similarly, for a fixed bulk modulus, increasing the length (aspect ratio) increases the tendency for failure on the high-pressure end. The reason is that, as the seal becomes longer, less of the applied load is supported by the low-pressure support, and more of it is transferred by shear to the rigid substrate. This shear stress is the source of failure on the high-pressure end.

\section{Conclusions}

A closed-form, approximate analytic solution has been derived for long, slightly compressible seals in both plane-strain and axisymmetric configurations. The solution reveals a key compressibility parameter, $\alpha \equiv \sqrt{3 \mu/\lambda} L/H$, proportional to length and inversely proportional to bulk modulus. As the shear modulus increases relative to the bulk modulus, or the length increases relative to the thickness, the seal becomes more compressible. The corresponding volumetric shrinkage means that there is deformation on the high-pressure end, and the corresponding shear transmits load to the rigid substrate rather than transmitting it to the mechanical support at the low-pressure end of the seal. In summary, as the seal becomes more compliant, the load transmitted to the rigid substrate increases and the load transmitted to the low-pressure support decreases. The approximate analytic solution agrees with finite element computations for finite deformation neo-Hookean materials.

Additionally, it was shown that an approximate Saint-Venant type solution, valid only away from the boundary layers, could be used to correlate energy release rates for cracks inside the boundary layers. The Saint-Venant solution provides the far-field loading stresses for the cracks, which are required to determine the amplitudes of the stress intensity factors and energy release rates. In functional form, we could write
\begin{equation}
	\mathcal{G} = \mathcal{G} \left( \boldsymbol{\sigma}^{\infty}; \mu, \mathrm{Geometry} \right)
\end{equation}
where $\boldsymbol{\sigma}^{\infty}$ is the far-field stress tensor provided by the Saint-Venant solution, and the effect of the remaining parameters are calculated via finite element analysis. In this case we are aided by the fact that the material is very nearly incompressible, which means the effect of shear modulus, $\mu$, and bulk modulus $K \approx \lambda$ are decoupled. The finite element simulations can be run for a particular choice of $\mu$, and the results can then be used for a wide variety of values of $K$ or $\lambda$. We expect this would not be the case if the Poisson ratio were allowed to vary of a large range because the energy release rate would then have a functional dependence of the form: $\mathcal{G} = \mathcal{G} \left( \boldsymbol{\sigma}^{\infty}; \mu, \nu, \mathrm{Geometry} \right)$.

The correlation between the analytic solution and energy release rates can be used to understand why making a seal longer does not necessarily equate to larger supported differential pressure in situations where friction between the seal and external sealing surface is small. The analytically computed phase diagram showing the region where failure will first occur on the high-pressure end and the region where failure will first occur on the low-pressure end agrees well with finite element calculations. This phase diagram can be used to explore how changing the bulk modulus or changing the aspect ratio will affect the location of failure. In practice, for a given elastomeric compound, the ratio of bulk modulus to shear modulus can be adjusted to a limited extent by varying crosslink density, and thereby varying shear modulus. However, it is much easier to vary the length of the seal in engineering designs, and this type of phase diagram could be used as a guide when choosing an appropriate length of seal for given material properties. In reality, it is not recommended that engineering designs be made solely based on this phase plot. Instead, this diagram illustrates the trade-off between length and compressibility, and should serve as a guide when screening preliminary seal designs. Detailed finite element modeling will undoubtedly be needed to account for more realistic conditions such as friction.

Although the current analysis only examined fracture initiation, the results suggest that a crack on the high-pressure end of a long, compressible seal will grow until the seal becomes effectively shorter and a critical energy release rate is reached on the low-pressure end. Conversely, for a very short, highly incompressible seal, the solution suggests that cracks will begin to grow on the low-pressure end of the seal until the seal is sufficiently compliant such that cracks on the high-pressure end begin to grow. In this way, it is not surprising that experiments have revealed damage on both ends of the seal regardless of its length. A fracture \emph{propagation} model, as opposed to the current examination of fracture initiation, is needed to show this result more conclusively as well as the breaking of axisymmetry leading to eventual seal leakage.


\begin{acknowledgments}
We appreciate valuable discussions with 
N. Wicks and E. B. Dussan V. and the suggestion by P. M. Reis to seek an axisymmetric solution. Roberta Mazzoli's assistance with high-performance computing for the Abaqus simulations is greatly appreciated.
We gratefully acknowledge financial support from Schlumberger-Doll Research.
\end{acknowledgments}

\section{Bibliography}
\bibliography{compressibility}

\appendix
\section{\label{Appendix: Cylindrical Coordinates} Axisymmetric Geometry}
\numberwithin{equation}{section}
\numberwithin{figure}{section}
\numberwithin{table}{section}
\setcounter{figure}{0}
\setcounter{table}{0}
The annular geometry is defined in Fig. \ref{Fig: Stacking Schematic3}, where the domain of the boundary value problem is $\mathbf{r}^{*} = \left( r^{*}, z^{*} \right)$ such that $R_{1} \le r^{*} \le R_{3}$ and $0 \le z^{*} \le L$. The displacement vector in this domain is
\begin{equation}
	\mathbf{u}^{*} \left( \mathbf{r}^{*} \right) = u_{r}^{*} \left( r^{*}, z^{*} \right) \hat{\mathbf{e}}_{r} + u_{z}^{*} \left( r^{*}, z^{*} \right) \hat{\mathbf{e}}_{z}
\end{equation}
The field equations and boundary conditions are analogous to those presented above for the plane strain case. We nondimensionalize the independent variables as
\begin{equation}
	r^{*} = R_{1} + H r
\end{equation}
where 
\begin{math}
	H \equiv R_{3} - R_{1},
\end{math}
giving
\begin{math}
	0 \le r \le 1
\end{math}.
We define a nondimensional ratio of lengths
\begin{equation}
	\beta \equiv \frac{H}{R_{1}}
\end{equation}
which gives the importance of curvature. As $\beta$ becomes small, the effect of curvature decreases and, in the limit as $\beta \rightarrow 0$, the plane-strain case is recovered. Then the independent variables are nondimensionalized as
\begin{subequations}
	\begin{align}
		r^{*} = & R_{1} \left( 1 + \beta r \right) \\
		z^{*} = & L z
	\end{align}
\end{subequations}
The dependent displacement variables are nondimensionalized as
\begin{equation}
	u_{r}^{*} = U u_{r}, \ \ \ u_{z}^{*} = W u_{z}
\end{equation}
with
\begin{equation}
	U = \frac{p_{0} H}{K}, \ \ \ W = \frac{p_{0} L}{K}
\end{equation}		
Then the nondimensionalized equilibrium equations are
\begin{subequations}
	\begin{multline}
		\frac{\partial}{\partial r}  \left( 2 \frac{\mu}{\lambda} \frac{\partial u_{r}}{\partial r} + \mathrm{tr} \boldsymbol{\epsilon} \right)
		+ \frac{\mu}{\lambda} \frac{\partial }{\partial z} \left( \frac{H^{2}}{L^{2}} \frac{\partial u_{r}}{\partial z} + \frac{\partial u_{z}}{\partial r} \right)
		\\
		+ 2 \beta \frac{\mu}{\lambda} \frac{ \left( \frac{\partial u_{r}}{\partial r} - \frac{\beta u_{r}}{1 + \beta r} \right)}{ \left( 1 + \beta r \right)} = 0
	\end{multline}
	\begin{multline}
		\frac{\mu}{\lambda} \frac{\partial }{\partial r} \left( \frac{\partial u_{r}}{\partial z} + \frac{L^{2}}{H^{2}} \frac{\partial u_{z}}{\partial r} \right)
		+ \frac{\partial }{\partial z} \left( \frac{\mu}{\lambda} \frac{\partial u_{z}}{\partial z} + \mathrm{tr} \boldsymbol{\epsilon} \right)
		\\
		+ \beta \frac{\mu}{\lambda} \frac{ \left( \frac{\partial u_{r}}{\partial z} + \frac{L^{2}}{H^{2}} \frac{\partial u_{z}}{\partial r} \right)}{\left( 1 + \beta r \right)} = 0
	\end{multline}
\end{subequations}
We again ignore small terms of order $H/L \ll 1$, $\mu/\lambda \ll 1$ while retaining terms of leading order $\left( \mu/\lambda \right) \left(L^{2}/H^{2} \right) \sim 1$ giving the leading-order governing equations
\begin{subequations}
	\begin{equation}
		\frac{\partial \left( \mathrm{tr} \boldsymbol{\epsilon} \right)}{\partial r} = -\frac{\partial p}{\partial r} = 0
	\end{equation}
	\begin{equation}
		\frac{\mu}{\lambda} \frac{L^{2}}{H^{2}} \frac{\partial^{2} u_{z}}{\partial r^{2}}
		+ \frac{\partial }{\partial z} \left(  \mathrm{tr} \boldsymbol{\epsilon} \right)
		\\
		+ \frac{ \beta \frac{\mu}{\lambda} \frac{L^{2}}{H^{2}} \frac{\partial u_{z}}{\partial r}}{\left( 1 + \beta r \right)} = 0
	\end{equation}
\end{subequations}
Upon integration and application of boundary conditions on the surfaces $r^{*} = R_{1}, R_{3}$, we arrive at the solution
\begin{tiny}
\begin{subequations}
	\begin{align}
		u_{r} \left( r , z \right) 
		= & p \left( z \right)
		\left\{-(2 + \beta) \beta^2 r (r-1) \left[4 + \beta \left( 2 + 4 r \right) + \beta ^2 r (r+1) \right]
		\right. \nonumber \\ & \left.
		+ 4 (1 + \beta)^2 \left[ (1 + \beta r)^2 (2 + \beta) \ln (1 + \beta r) 
		\right. \right. \nonumber \\ & \left. \left.
		- r (1 + \beta)^2 (2 + \beta r) \ln (1 + \beta) \right] \right\} /
\nonumber \\ &
	\left\{ 2 (1 + \beta r) \left[4 (1 + \beta)^4 \ln (1 + \beta)-\beta \left(4 + 14 \beta + 12 \beta^{2} + 3 \beta^3\right)\right] \right\}
		\\
		u_{z} \left( r, z \right)
		= &
		\frac{H^{2}}{L^{2}} \frac{\lambda}{\mu} \left[ \frac{\beta r \left( 2 + \beta r \right) 
		-2 \left( 1 + \beta \right)^{2} \ln \left( 1 + \beta r \right)}{4 \beta^{2}} \right] \frac{dp}{dz}
	\end{align}
	\label{Eqn: Analytic Solution for Displacements in Cylindrical Geometry}
\end{subequations}
\end{tiny}
with
\begin{equation}
	p \left( z \right) = D_{1} \cosh \left( \gamma z \right) + D_{2} \sinh \left( \gamma z \right)
\end{equation}
where
\begin{subequations}
	\begin{align}
		\gamma 
		\equiv &
		f ( \beta ) \sqrt{\frac{L^{2}}{H^{2}} \frac{\mu}{\lambda}}
	\\
		f( \beta ) 
		\equiv &
		\frac{2 \sqrt{2} \beta ^{3/2} \sqrt{\beta +2}}{ \sqrt{4 (\beta +1)^4 \ln \left(\beta +1\right)-\beta  \left(3 \beta ^3+12 \beta ^2+14 \beta +4\right)}}
	\end{align}
\end{subequations}
where the Taylor series expansion of $f \left( \beta \right)$ for values of $\beta$ near zero in the plane-strain limit is
\begin{equation}
	f \left( \beta \right) = \sqrt{3} \left( 1 - \frac{1}{4} \beta + \frac{23}{160} \beta^{2} + \mathcal{O} \left( \beta^{3} \right) \right)
\end{equation}
which can be clearly seen to have the limiting value of $f \left( \beta \right) \rightarrow \sqrt{3}$ and $\gamma \rightarrow \alpha$ in the plane-strain limit of $\beta \rightarrow 0$. 

As in the plane strain case, there are two constants of integration and four axial boundary conditions. We choose to satisfy the boundary conditions normal to the surfaces at $z^{*} = 0, L$ in the integral sense.
\begin{subequations}
	\begin{multline}
	\int_{R_{1}}^{R_{3}}{\sigma_{zz}^{*} 2 \pi r^{*} dr^{*}} = - \pi \left( R_{3}^{2} - R_{1}^{2} \right) p_{0} \ \ \mathrm{on} \ \ z^{*} = 0
	\\ \Rightarrow
	2 R_{1} H \int_{0}^{1}{\sigma_{zz} \left( r, z = 0 \right) \left( 1 + \beta r \right) dr} = -\left( R_{3}^{2} - R_{1}^{2} \right)
	\end{multline}

	\begin{multline}
	\int_{R_{1}}^{R_{3}}{2 \pi r^{*} \left( \sigma_{zz}^{*} + k^{*} u_{z}^{*} \right) dr^{*}} = 0 \ \ \mathrm{on} \ \ z^{*} = L
	\\ \Rightarrow
	\int_{0}^{1}{\left[ \sigma_{zz} \left( r, z = 1 \right) + k u_{z} \left( r, z = 1 \right) \right] \left( 1 + \beta r \right) dr} = 0
	\end{multline}
\end{subequations}
where $k = \frac{k^{*} L}{K}$ is the nondimensional low-pressure stiffness. Together, these boundary conditions specify the values of the integration constants as
\begin{subequations}
	\begin{align}
		D_{1} = & \frac{K}{\lambda + 2 \mu}
	\\ 
		D_{2} = & -\frac{\cosh \gamma + \frac{k K}{\gamma \left( \lambda + 2 \mu \right) } \sinh \gamma}
		{\frac{\lambda + 2 \mu}{K} \sinh \gamma + \frac{k}{\gamma} \cosh \gamma}
	\end{align}
\end{subequations}

\begin{figure}[h!t]
	\centerline{\includegraphics[width=\columnwidth]{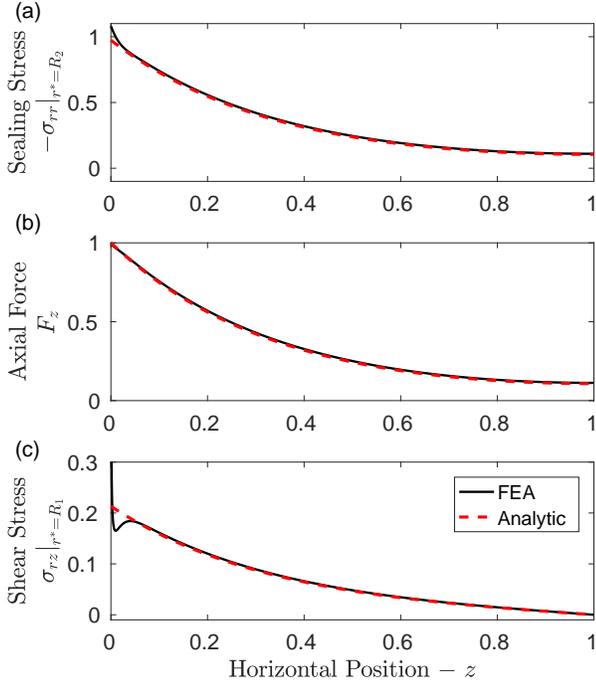}}
		\caption{Comparison between finite element prediction and approximate analytic solution for axisymmetric geometry. Analog to plane strain comparison in Fig. \ref{Fig: Stress vs Axial Position}. The horizontal axis is stretched to show the detailed differences between the analytic and computational solutions at the ends. Finite element results for a neo-Hookean material undergoing finite deformations. The relevant parameters for the simulation are given in Table \ref{Table: Params for Axisymmetric Model}. (a) Sealing stress between upper surface of seal and top rigid surface at $r^{*} = R_{3}$. (b) Axial force, $F_{z}$, along the length of the seal. (c) Shear stress at interface where seal is bonded to inner rigid surface at $r^{*} = R_{1}$.}
		\label{Fig: Stress vs Axial Position - Axisymmetric}
	\end{figure}

Fig. \ref{Fig: Stress vs Axial Position - Axisymmetric} provides a comparison between finite element simulations and approximate analytic solution results for profiles of key stress measures along the length of the seal. The input parameters for the finite element simulations are given in Table \ref{Table: Params for Axisymmetric Model}, and are for a value of $\beta = 1$ indicating the effect of curvature is leading order. The results are analogous to those in Fig. \ref{Fig: Stress vs Axial Position}, which were for the plane strain case. As in the plane-strain case, the approximate analytic results agree exceptionally well with the finite element results except in the boundary layer regions near the ends of the seal for $z^{*} < H$ and $L - H < z^{*} < L$.

	\begin{table}[h!]
		\caption{Parameter values for validation of axisymmetric analytic solution with finite element simulation.}
		\label{Table: Params for Axisymmetric Model}
		\begin{tabular}{|l|l|}
			\hline
			\textbf{Parameter} & \textbf{Value} \\ \hline
			Outer Radius: $R_{3} / R_{1}$	&	2 \\ \hline
			Thickness: $H / R_{1}$	&	1 \\ \hline
			Length: $L / R_{1}$	&	20	 \\	\hline
			LP Support Height: $h / H$	&	1 \\ \hline
			LP Support Stiffness: $k^{*}$	&	$\infty$	\\ \hline
			Curvature: $\beta$	&	1	\\ \hline
			Bulk Modulus: $K / \mu$	&	100	\\ \hline
			Applied Load: $p_{0} / \mu$	&	0.5	\\ \hline
			Solution Parameter: $f \left( \beta \right)$	&	1.453 \\ \hline
			Length Scaling (Plane Strain): $\alpha$	&	3.476	\\ \hline
			Length Scaling (Axisymmetric): $\gamma$	&	2.917	\\ \hline
		\end{tabular}
	\end{table}

\end{document}


\maketitle
\tableofcontents

\renewcommand{\thefigure}{SM-\arabic{figure}}
\renewcommand{\theequation}{SM-\arabic{equation}}
\renewcommand{\thetable}{SM-\arabic{table}}
\section{Experimental Investigation of Seal Length}

This work was motivated by a small set of experiments that investigated the effect of seal length on maximum differential pressure supported by a seal. Although these experiments were not exhaustive in the investigation of geometry and material property parameters, the results for this set of experiments reveal that, for the particular conditions of these experiments, stacking two sealing elements in series does not affect the maximum differential pressure supported by the seal system. It is true that these tests were performed in steel pressure vessels with relatively smooth walls, so the effect of friction was minimized. However, steel pressure vessels or steel casing are typical for validation testing of a novel packer design prior to deployment in oil and gas wells. The lack of dependence of critical differential pressure on seal length in these experiments suggests that previous understanding about scaling of maximum differential pressure with seal length may not be valid in all cases. It is this invalidity, for this specific set of experiments, that motivates our investigation of the combined effects of length and compressibility on performance of long elastomeric seals.

A brief overview of the geometry and key results is given in Section II of the main article. Here we expand upon the description. The fundamental modular sealing element consists of an annular body of carbon black-filled EPDM-based rubber bonded to a thin annular steel shell. The outer diameter of the rubber element is $D = 4.61 \ \mathrm{cm}$ and the diameter of the interface between the inner surface of rubber and outer surface of the shell is $D_{1} = 3.49 \ \mathrm{cm}$. The length of rubber and steel shell is $L = 10.2 \ \mathrm{cm}$. In the experiments either a single element or two elements with nothing between them were slid onto a solid steel mandrel of outer diameter $D = 3.06 \ \mathrm{cm}$. Metal support rings with outer diameter $D_{2} = 4.61 \ \mathrm{cm}$ were affixed to the steel mandrel on either end of the seal element(s) with either set screws or shaft collars. The mandrel with seal element(s) was inserted into a pressure vessel. The pressure vessel, mounted vertically to eliminate bubbles, was filled from the bottom with a low-viscosity mineral oil (Penreco LVT-200), which is a standard oil blend for packer testing in industry. The pressure vessel was heated to $T = 82^{\circ}$C using electric resitance heating bands (Briskheat BSAT201006 and BSAT201008), and temperature was controlled using a simple thermostat (Micromatic E54 Brewer's Edge\textsuperscript{\textregistered} Controller II). The seals were swollen at elevated temperature for two or three weeks, as denoted in Table SM-\ref{Table: Seal Stacking}.

At the conclusion of swelling, the seals were subjected to differential pressure applied using an Teledyne Isco Model 260D syringe pump containing the same LVT 200 mineral oil. The pump was programmed to continuously ramp pressure at a rate of $1.15 \ \mathrm{kPa/s}$ ($10$ psi/min). The low-pressure end of the pressure vessel was open to atmospheric pressure; therefore, the gauge pressure supplied by the pump was equivalent to the differential pressure across the seal. Pump pressure was increased until there was a precipitous drop in pressure and a sharp spike in flow rate. The maximum pump pressure was denoted as the critical differential pressure.

\begin{figure}[h!t]
	\centerline{\includegraphics[scale=0.5]{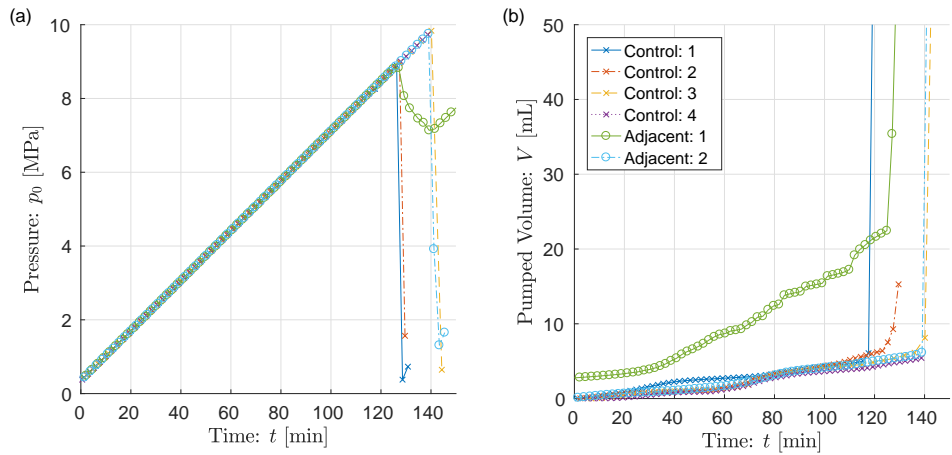}}
	\caption{Syringe pump data for experiments investigating effect of seal length on critical differential pressure. (a) Pressure, as measured at the syringe pump, as a function of time. (b) Displaced volume by syringe pump. The continuous rise of pumped volume is due in part to compliance in the system, but is also partially due to extrusion of the seal material on the low-pressure end of the seal prior to seal leakage.}
	\label{Fig: Experimental Data - Stacking}
\end{figure}

Four replications of a control configuration and two replications of the trial configuration were tested. The swelling duration, configuration, and maximum differential pressure are reported in Table SM-\ref{Table: Seal Stacking}. Pressure and pumped volume versus time, as output from the pump, for each of the six tests are plotted in Fig. \ref{Fig: Experimental Data - Stacking}.

\begin{table}[h!]
	\label{Table: Seal Stacking}
	\caption{Summary of experimental results investigating effect of stacking of seal elements.}
	\begin{center}
	\begin{tabular}{|l|c|c|c|}
		\hline
		Test	&	Swell Duration	&	Critical $\Delta P$ & Critical $\Delta P$ \\
				&	[days]			&	[MPa]				&	[psi]	\\ \hline
		Control 1	&	21			&	9.43					&	1370		\\ \hline	
		Control 2	&	21			&	9.16					&	1330		\\ \hline	
		Control 3	&	14			&	9.99					&	1450		\\ \hline	
		Control 4	&	14			&	10.0					&	1450		\\ \hline	
		Adjacent 1	&	23			&	9.06					&	1310		\\ \hline	
		Adjacent 2	&	21			&	10.0					&	1450		\\ \hline	
	\end{tabular}
	\end{center}
\end{table}

\section{Calibration of Low-Pressure Support Stiffness, $k^{*}$}
In the analytic solution presented in the paper, the low-pressure support at $x^{*} = L$ ($z^{*} = L$ in axisymmetric geometry) is parameterized as a linear spring with stiffness coefficient $k^{*}$ such that the applied stress is related to the deformation at the low-pressure surface through
\begin{subequations}
\begin{equation}
	\overline{\sigma}_{11}^{*} + k^{*} \overline{u}^{*} = 0 \ \mathrm{ on } \ x^{*} = L
\end{equation}
for the plane-strain case and correspondingly
\begin{equation}
	\overline{\sigma}_{zz}^{*} + k^{*} \overline{u}_{z}^{*} = 0 \ \mathrm{ on } \ z^{*} = L 
\end{equation}
\end{subequations}
for the axisymmetric configuration, where the overline denotes the average value over the cross-section. The entire analytic solution has only this single adjustable parameter, which is not solved for explicitly as part of the model. However, a single finite element simulation for a given geometry $\left( h, H, r_{c} \right)$ and shear modulus $\mu$ can be used to determine $k^{*}$. \footnote{Note that we have observed no dependence of $k^{*}$ on bulk modulus for these nearly incompressible materials. However, the precise dependence of $k^{*}$ on material properties is unknown.} Then this value can be used as the length and bulk modulus of the seal are varied.

\begin{figure}[h!t]
	\centerline{\includegraphics[scale=0.50]{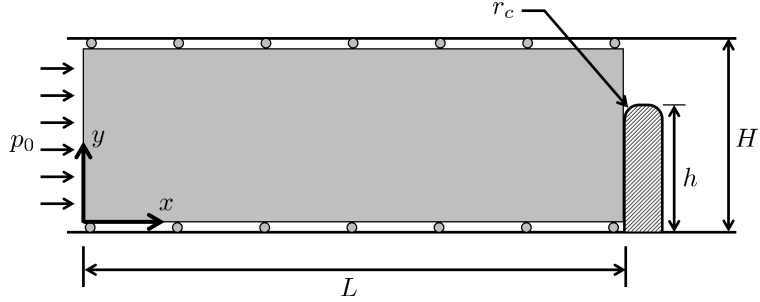}}
	\caption{System schematic for finite element simulations to empirically determine the stiffness of the low-pressure support, $k^{*}$.}
	\label{Fig: Geometry - LP Support Stiffness}
\end{figure}

In order to determine this effective stiffness of the low-pressure support, we conduct a single finite element analysis of a short elastomeric seal, similar to that sketched in Fig. \ref{Fig: Geometry - LP Support Stiffness}, having a length, $L$, that is on the order of four times the thickness, $H$, of the seal. (The factor of four is empirically found to be sufficiently far from the low-pressure support such that the deformation is nearly uniform on the high-pressure surface.) 

The objective of this calibration is to quantify the relation between stress and deformation at the low-pressure support. In order to separate the effect of shear on the lateral surfaces from the effect of mechanical support on the low-pressure surface, we prescribe shear-free boundary conditions on both the lower and upper surfaces at $y^{*} = 0$ and $y^{*} = H$. (In the corresponding axisymmetric configuration, shear-free boundaries at the inner radius, $r^{*} = R_{1}$, and outer radius, $r^{*} = R_{3}$, are prescribed.) The input to the finite element simulation is the applied load, $p_{0}$, and the output given is the resulting nodal displacements on the high-pressure end at $x^{*} = 0$ ($z^{*} = 0$ in axisymmetric geometry). To calculate an effective stiffness in plane strain, we evaluate the average displacement of the high-pressure end at $x^{*} = 0$ by numerically integrating the nodal displacements.
\begin{equation}
	\overline{u}_{0}^{*} \equiv \frac{1}{H} \int_{0}^{H}{u^{*} \left( x = 0, y \right) dy}
\end{equation}
The total displacement on the high-pressure end is the result of both the decrease in volume of the body due to compression as well as isochoric extrusion into the low-pressure gap. The stiffness, $k^{*}$, gives a relation between the part of displacement due to isochoric extrusion. We therefore partition the total displacement on the high-pressure end into that which is caused by bulk compression, and that which is due to isochoric extrusion.
\begin{equation}
	\overline{u}_{0}^{*} = \overline{u}_{\mathrm{0,bulk}}^{*} + \overline{u}_{\mathrm{0,extrusion}}^{*}
\end{equation}
The bulk deformation is the deformation that would result if the low-pressure end were infinitely stiff, and there were only uniform uniaxial strain due to volume reduction. Since the bulk modulus i slarge and the volume change small, we expect an infinitesimal deformation solution for uniaxial compression to give reasonable estimate for $\overline{u}_{\mathrm{0,bulk}}^{*}$. The infinitesimal displacement solution for uniaxial strain gives
\begin{equation}
	\sigma_{11}^{*} = \left( 2 \mu + \lambda \right) \epsilon_{11}^{*} = \frac{E \left( 1 - \nu \right)}{\left( 1 + \nu \right) \left( 1 - 2 \nu \right)} \epsilon_{11}^{*}
\end{equation}
For uniform applied stress $\sigma_{11}^{*} = -p_{0}$, integrating the infinitesimal strain $\epsilon_{11}^{*} = \partial u^{*}/\partial x^{*}$ over the length of the body gives
\begin{equation}
	\overline{u}_{0,\mathrm{bulk}}^{*} = u^{*} \left( x^{*} = 0, y \right) = \frac{p_{0} L}{2 \mu + \lambda} = p_{0} L \frac{\left( 1 + \nu \right) \left( 1 - 2 \nu \right)}{E \left( 1 - \nu \right)}
\end{equation}
where $\mu$ and $\lambda$ are the small-strain Lam\'{e} moduli (tangent moduli at zero strain), and we have prescribed zero displacement, $u^{*} = 0$, at the low-pressure end of the seal, $x^{*} = L$. Finally, the stiffness is evaluated as
\begin{equation}
	k^{*} = \frac{p_{0}}{\overline{u}_{0,\mathrm{extrusion}}^{*}} = \frac{p_{0}}{\overline{u}_{0}^{*} - \overline{u}_{0,\mathrm{bulk}}^{*}}
	\label{Eqn: k-star}
\end{equation}

\begin{figure}[h!t]
	\centerline{
		\includegraphics[width=0.5\textwidth]{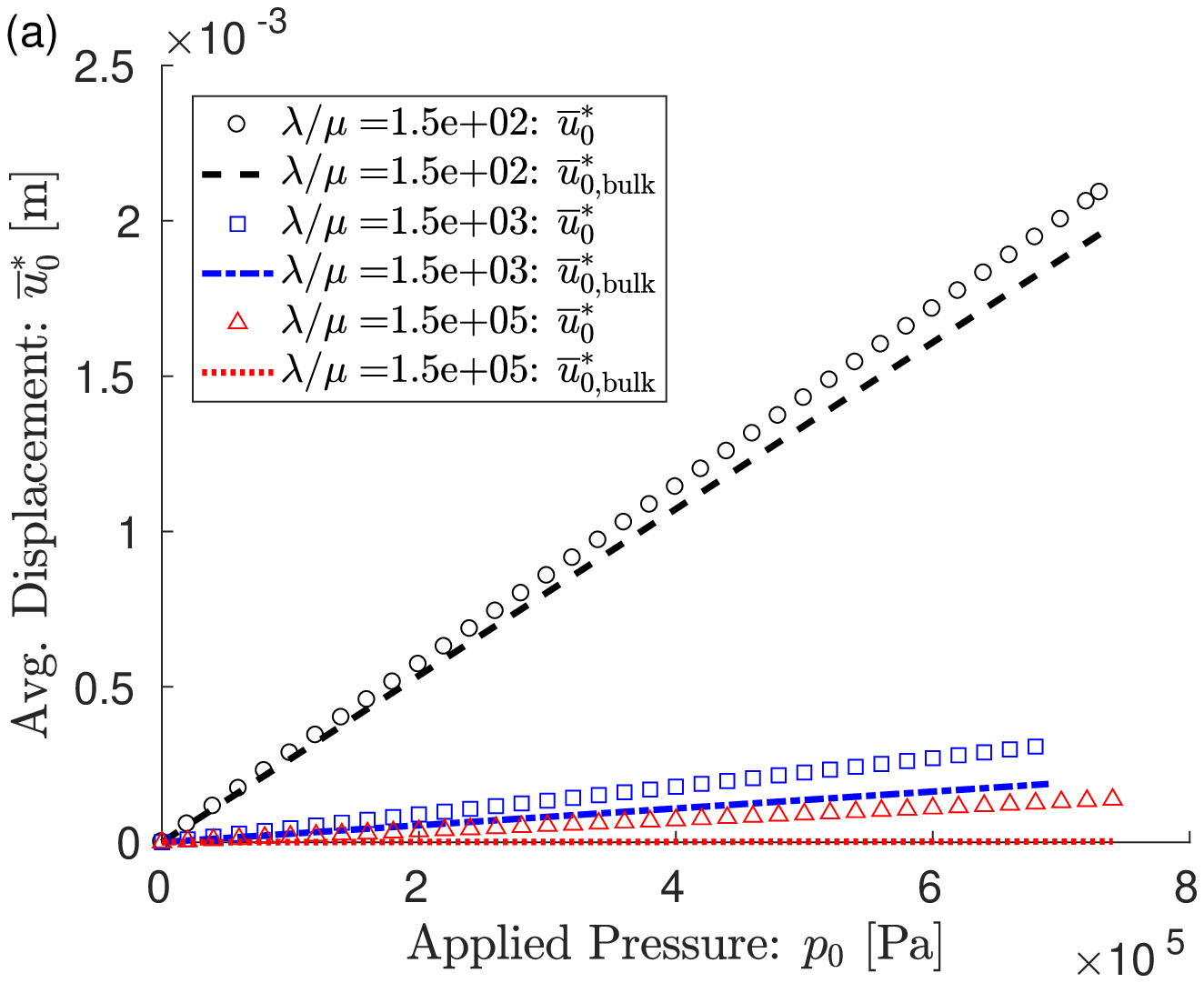}
		\includegraphics[width=0.5\textwidth]{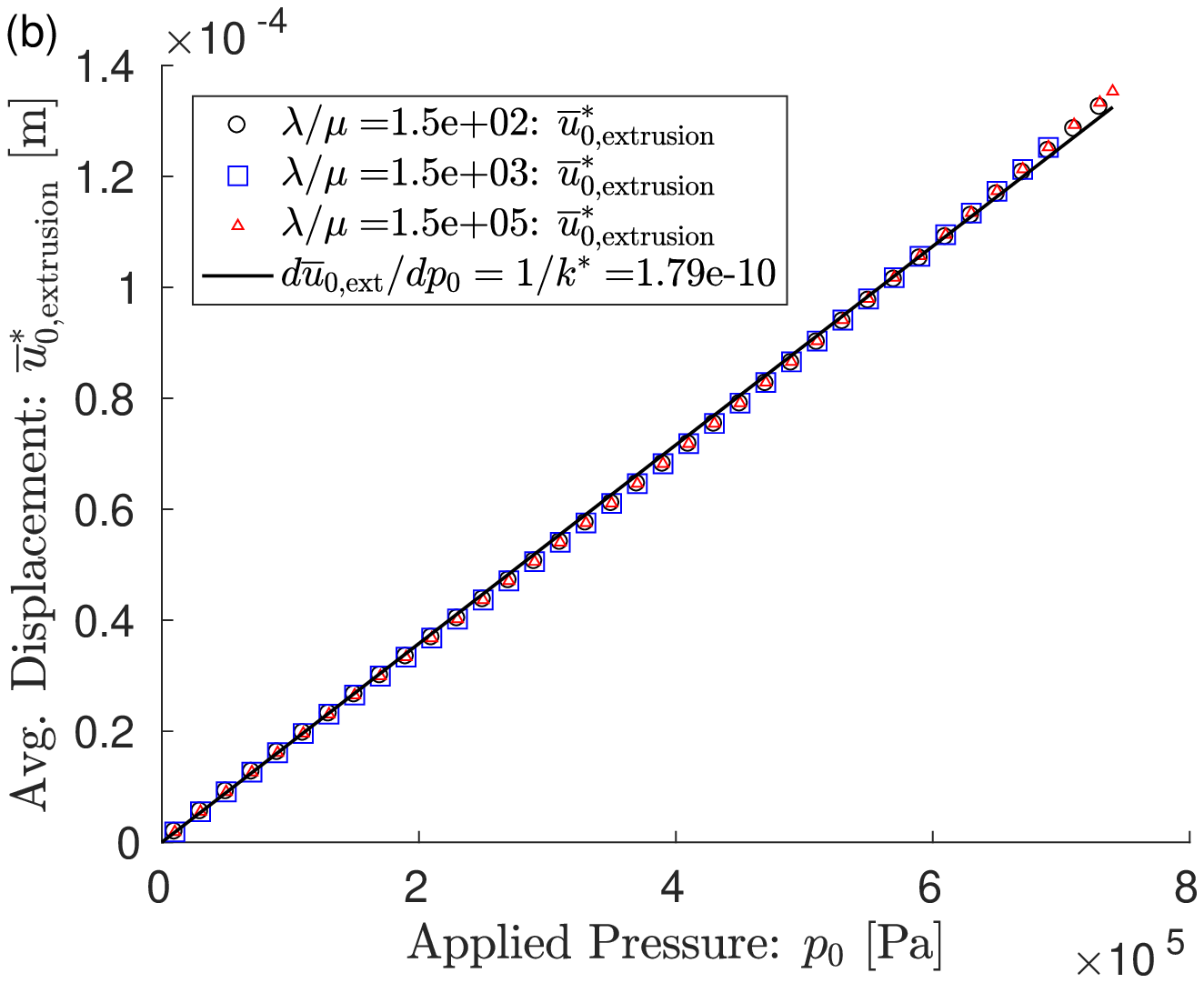}
	}
	\caption{Results for calibration of low-pressure support filling 90 percent of the gap, $H$, between inner and outer rigid surfaces. Parameter values: $\mu = 613 \ \mathrm{kPa}$, $H = 2.5 \ \mathrm{cm}$ (a) Average total displacement, $\overline{u}_{\mathrm{0}}^{*}$ and uniform uniaxial displacement, $\overline{u}_{\mathrm{0,bulk}}^{*}$. (b) Isochoric extrusion displacement, $\overline{u}_{\mathrm{0,extrusion}}^{*} = \overline{u}_{\mathrm{0}}^{*} - \overline{u}_{\mathrm{0,bulk}}^{*}$}
	\label{Fig: LP Support Stiffness}
\end{figure}

Fig. \ref{Fig: LP Support Stiffness} shows results for calculation of the low-pressure stiffness for a seal of thickness $H = 2.5 \ \mathrm{cm}$, neo-Hookean material with small-strain shear modulus $\mu = 613 \ \mathrm{kPa}$, low-pressure support filling 90 percent of the seal height and a corner radius of $r_{c} = 0.188 \ \mathrm{mm}$. The results are shown for three significantly different small-strain bulk moduli in order to demonstrate that the results of support stiffness, $k^{*}$, are insensitive to bulk modulus. Despite the vastly different total average displacement, $\overline{u}_{0}^{*}$, for the three simulations, subtracting the bulk displacement, $\overline{u}_{\mathrm{0,bulk}}^{*}$, yields an isochoric deformation, $\overline{u}_{\mathrm{0,extrusion}}^{*}$, that is insensitive to bulk modulus.

\begin{figure}[h!t]
	\centerline{\includegraphics[trim={1.0cm 0.0cm 2.0cm 0.4cm},clip,width=\textwidth]{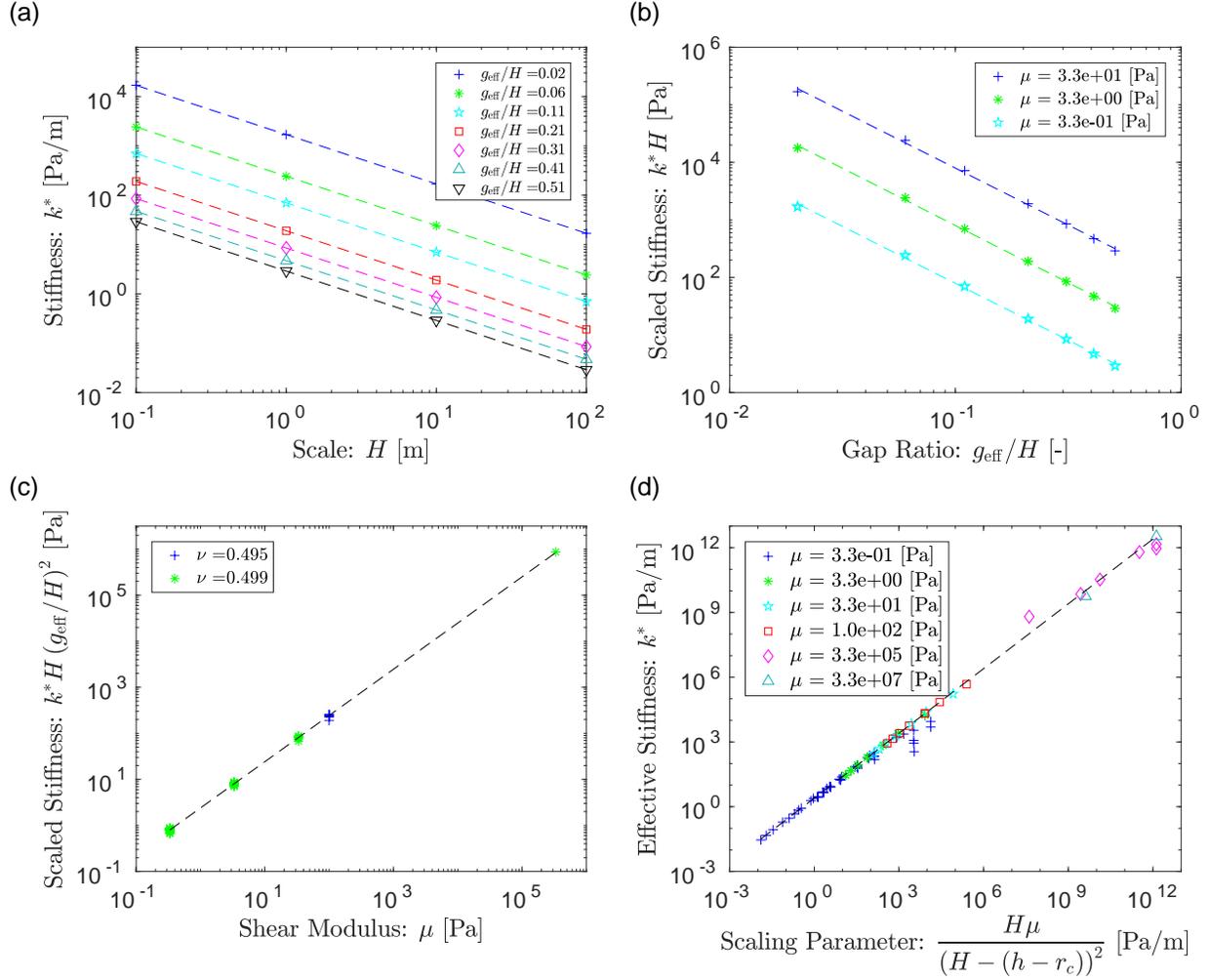}}
	\caption{Effective stiffness of low-pressure support for infinitesimal deformation linear elasticity under plane strain. (a) Stiffness versus seal thickness, $H$. Parameters: $E = 1 \mathrm{\ Pa}$, $\nu = 0.499$ and $r_{c} / H = 0.01$.
	(b) Stiffness versus gap ratio, after scaling out seal thickness. Parameters: $\nu = 0.499$ and $r_{c}/H = 0.01$.
	(c) Stiffness versus shear modulus after scaling out the geometry. Parameters: $r_{c}/H = 0.01$.
	(d) Stiffness versus material and geometric parameters.}
	\label{Fig: Stiffness vs Geometry}
\end{figure}

In the preceding finite element calculations, we considered finite deformation of a neo-Hookean material, which is complicated because the contact condition on the low-pressure end changes as the material is elastically extruded into the gap. For the case of finite deformation, we have made no attempt to map the stiffness of the support as a function of the geometry and material parameters. However, we have attempted to do so, via finite element analysis, for the linear limit (which can likely alternately be solved analytically, perhaps using complex analysis techniques of Muskhelishvili.) For infinitesimal linear elastic deformations, we empirically find the low-pressure support stiffness for plane strain to be of the form
\begin{equation}
	k^{*} = 2.5 \frac{H \mu}{\left( H - \left( h - r_{c} \right) \right)^{2}}
	= 2.5 \frac{H \mu}{ g_{\mathrm{eff}}^{2}}
\end{equation}
where the prefactor $2.5$ is determined from curve fitting and we define the effective gap, $g_{\mathrm{eff}}$ to be the height between the outer sealing surface and the lower corner of the fillet on the low-pressure support. Plots of finite element simulations versus this empirical result are shown in Fig. \ref{Fig: Stiffness vs Geometry}. Note that we have considered only a very narrow range of Poisson ratio near the incompressible limit of $\nu \rightarrow 1/2$, which is of interest for elastomers. We do not know how Poisson ratio will affect the effective stiffness if allowed to deviate far from $\nu \approx 1/2$.

Note that the behavior of the effective stiffness for finite deformations is complicated by the fact that the seal deforms around the corner radius. Therefore, the effective gap, $g_{\mathrm{eff}}$, may be better approximated by $H - h$ rather than $H - \left( h - r_{c} \right)$ in the infinitesimal deformation case. For finite deformation cases, the effective stiffness for a material undergoing infinitesimal deformation can be used as a starting point, and a single finite element analysis can determine the more precise value for a given geometry and shear modulus.

\section{Finite Element Simulations for Crack Initiation}

\begin{figure}[h!t]
	\centerline{\includegraphics[scale=0.50]{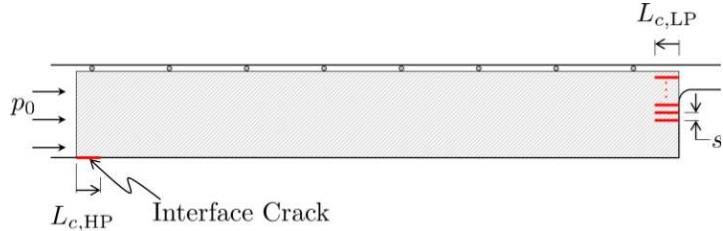}}
	\caption{Schematic showing orientation of cracks for evaluation of energy release rate, $\mathcal{G}$, for characterization of tendency for fracture initiation.}
	\label{Fig: Schematic - Evaluation of J Integral}
\end{figure}

In the paper, the energy release rates for the growth of pre-existing cracks on both the high-pressure and low-pressure ends are discussed. Finite element simulations are used to evaluate these energy release rates for finite-deformation, neo-Hookean materials under plane strain using the CPE4H element type in the software package Abaqus/Standard. Fig. \ref{Fig: Schematic - Evaluation of J Integral} is a schematic showing the seal under plane strain deformation, along with the location of cracks introduced into the mesh for the purpose of evaluation of the energy release rates. On the high-pressure end, a single interfacial crack of length $L_{c,\mathrm{HP}} = 10 \ \mu \mathrm{m}$ is introduced at the interface between the elastic body and the rigid substrate. We impose boundary conditions on this crack of zero shear traction and zero normal displacement. The choice of these is subjective. The choice of zero normal displacement was made because it was presumed that, in realistic applications, the seal will be initially pre-compressed in the direction normal to the inner and outer sealing surfaces. This pre-compression will be caused by swelling for the case of swell packer seals or mechanical compression upon assembly or due to application of axial load, as in the case of mechanical packers. Regardless, it is presumed that, in order for a seal to be maintained, it is necessary that the compressive traction normal to the sealing surface be larger than the applied hydrostatic load from fluid pressure. Therefore, we presume this crack does not open in Mode I fracture.

The choice of no shear in the tangential direction is arbitrary. There will certainly be friction in realistic situations. However, friction has been neglected throughout the analytic model at the other surfaces, and we therefore neglect it here. In more applied finite element modeling, this assumption can easily be relaxed.

On the low-pressure end, we introduce a sequence of $N_{\mathrm{cracks}} = 51$ small cracks, each of length $L_{c,\mathrm{LP}} = 10 \ \mu \mathrm{m}$, perpendicular to the undeformed low-pressure surface. These cracks are uniformly spaced in the direction parallel to the face of the low-pressure support with spacing $s = 20 \ \mu \mathrm{m}$. The reason for the introduction of such a large number of cracks is that in physical systems, we expect there to be small cracks throughout the material, and we generally do not know the geometry of these pre-existing defects. The reason we care about these is that, at some critical pressure, one of the cracks will begin to grow. If we knew exactly which crack would begin to grow, we could omit many of the other cracks because the stress concentrations around their crack tips are locally confined. However, for the finite deformation finite element analysis, we do not know \emph{a priori} which crack will experience the maximum energy release rate. For small loads, and small deformations, the crack experiencing the largest energy release rate is one that is located near the corner of the low-pressure support in the undeformed configuration. However, as pressure is increased, this particular crack elastically extrudes into the extrusion gap, and a different crack experiences a larger energy release rate than the first. This transfer of peak energy release rate from one crack to another continues as pressure and elastic deformation increase. Because of this finite deformation, it is not a single material point (single crack) which always experiences the maximum energy release rate. Therefore, in order to calculate the tendency, at any given pressure, for crack growth on the low-pressure end, the maximum value over all 51 cracks is taken for each value of applied pressure.

\begin{equation}
	\mathcal{G}_{\mathrm{LP}}\left( p_{0} \right) = \max_{i}{\left(\mathcal{G}_{\mathrm{LP},i} \left( p_{0} \right)\right)} \ \ \left( i = 1, 2, \cdots, N_{\mathrm{cracks}} \right)
\end{equation}

\begin{figure}[h!t]
	\centerline{
		\includegraphics[width=0.5\textwidth]{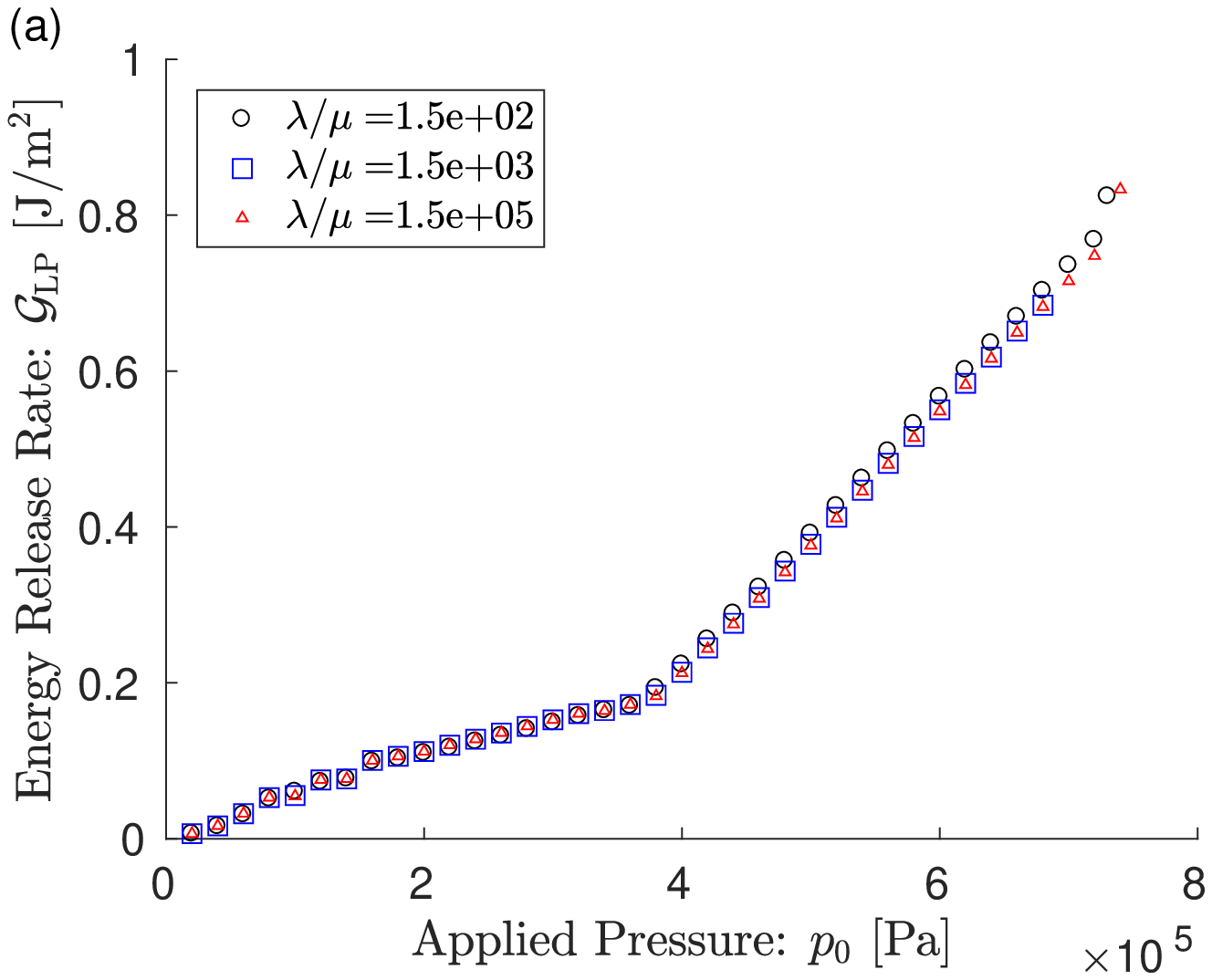}
		\includegraphics[width=0.5\textwidth]{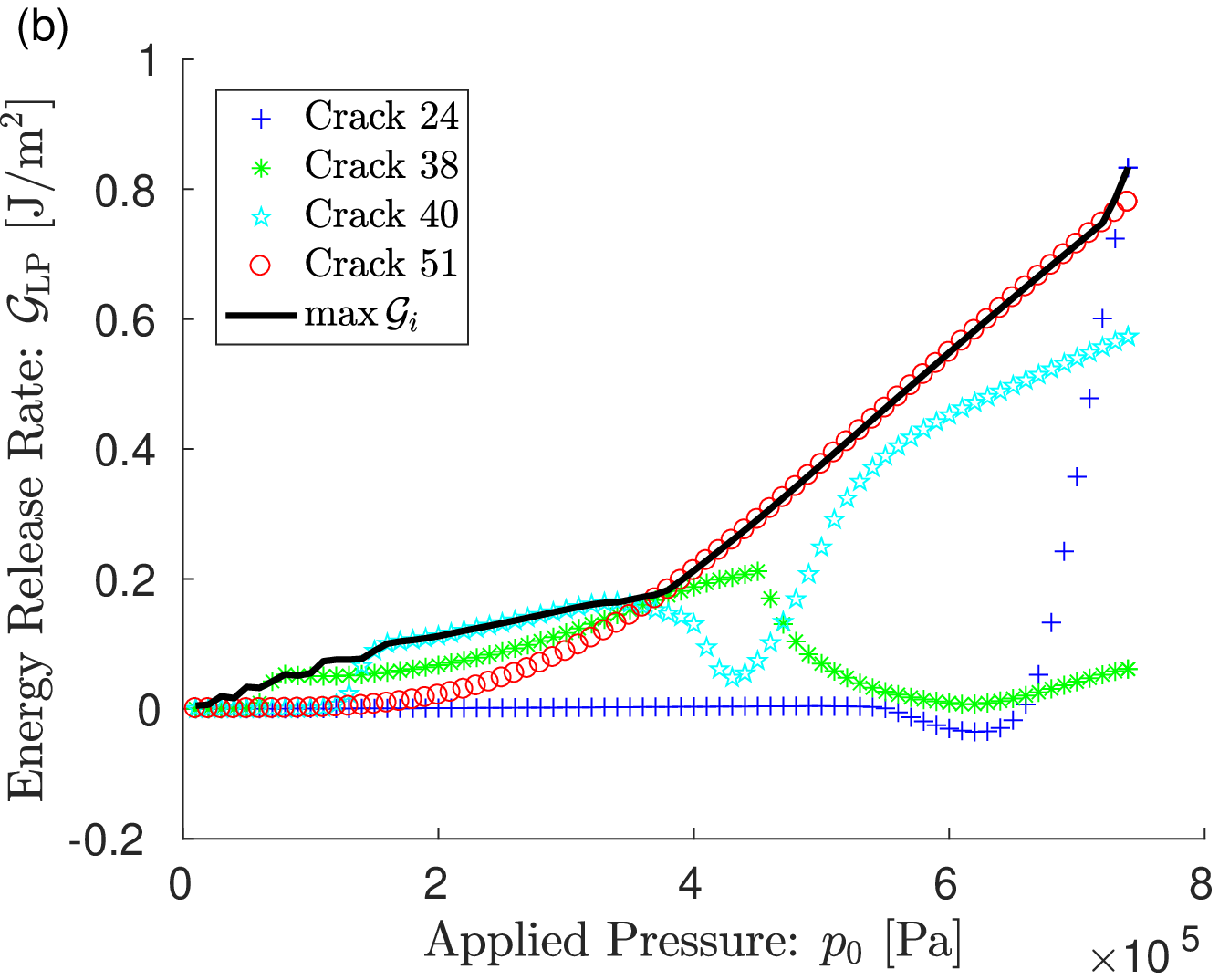}
	}
	\caption{Energy release rate for growth of cracks on low-pressure end, for shear-free inner and outer sealing surfaces. (a) Effect of bulk modulus on energy release rates showing $\mathcal{G}_{\mathrm{LP}}$ is relatively insensitive to $\lambda / \mu$. (b) Energy release rate for several individual cracks, $\mathcal{G}_{\mathrm{LP},i}$, on low-pressure end as a function of applied pressure for $\lambda / \mu = 1.5 \times 10^5$. $\mathcal{G}_{\mathrm{LP}}$ is the locus of maximum values over all the cracks for each value of applied pressure.}
	\label{Fig: LP ERR - No Shear}
\end{figure}

To illustrate the effects of multiple cracks on energy release rate at the low-pressure end, we again consider the system used to calculate low-pressure stiffness, $k^{*}$, as illustrated in Fig. \ref{Fig: Geometry - LP Support Stiffness}. Recall this system is shear-free on the inner and outer sealing surfaces, which isolates the low-pressure support effect from the shear lag effect. Fig. \ref{Fig: LP ERR - No Shear}.a shows the energy release rate, $\mathcal{G}_{\mathrm{LP}}$, on the low-pressure end, as a function of applied pressure for two different ratios of elastic moduli. The results show that the energy release rate on the low-pressure end is insensitive to bulk modulus provided shear modulus is held fixed.

Fig. \ref{Fig: LP ERR - No Shear}.b shows the energy release rates for several individual cracks, $\mathcal{G}_{\mathrm{LP},i}$ for a ratio of small-strain elastic moduli of $\lambda / \mu = 1.5 \times 10^{5}$. The results show that, for very small applied loads and small deformations, crack numbers 37, 38, 39 and 40 yield the largest energy release rates. As pressure is increased past approximately $p_{0} = 400 \mathrm{\ kPa}$, crack 51 yields the largest energy release rates. The reason for this is shielding from other cracks. Crack 51 is the top-most crack in the array of cracks, and therefore has only neighboring cracks below it. Therefore, it receives only half the shielding of the other cracks, and its energy release rate is larger, over a range of pressures, than cracks that are more shielded. Finally, at the highest pressures, cracks 24 and 25 yield the largest energy release rates. This transfer among cracks \{37-40\}, 51 and \{24-25\} is due to finite deformation. Together, the finite deformation and contact make this local energy release problem nonlinear and contribute to the peculiar shape of the resulting maximum energy release rate curve.

In the corresponding article, the effect of seal length and compressibility was investigated for two different low pressure support geometries. Figures 5 and 6 are finite element results for the case when the extrusion gap is 25 percent of the channel thickness, meaning the low-pressure support is relatively compliant and extrusion is facilitated. The geometry for these plots is given in Table SM-2.

\begin{table}[h!t]
	\label{Table: Larger Gap}
	\caption{Plane strain geometry -- 25 percent gap. (Figs. 5 and 6)}
	\begin{center}
	\begin{tabular}{|l|c|}
		\hline
		\textbf{Parameter}	&	\textbf{Value}	\\
					&	[m]	\\ \hline
		$h$			&	0.01875		\\ \hline
		$H$			&	0.025		\\ \hline
		$r_{c}$		&	0.000188		\\ \hline
		$L_{c,\mathrm{HP}}$		&	$1.0\times 10^{-5}$	\\ \hline
		$L_{c,\mathrm{LP}}$		&	$1.0\times 10^{-5}$	\\ \hline
		$s$			&	$2.0 \times 10^{-5}$	\\ \hline
		Mesh Size	&	\multirow{2}{*}{$1.0 \times 10^{-6}$} \\
		(Locally Refined) & \\ \hline
	\end{tabular}
	\end{center}
\end{table}

The alternate configuration is a low-pressure support geometry where the extrusion gap is ten percent of the channel width, and the details of this geometry are given in Table SM-3. This geometry provides a much larger low-pressure support stiffness, $k^{*}$, and the values from these simulations correspond to Fig. 7 in the journal article.

\begin{table}[h!t]
	\label{Table: Smaller Gap}
	\caption{Plane strain geometry  -- 10 percent gap. (Fig. 7)}
	\begin{center}
	\begin{tabular}{|l|c|}
		\hline
		\textbf{Parameter}	&	\textbf{Value}	\\
					&	[m]	\\ \hline
		$h$			&	0.0225		\\ \hline
		$H$			&	0.025		\\ \hline
		$r_{c}$		&	0.000188		\\ \hline
		$L_{c,\mathrm{HP}}$		&	$1.0\times 10^{-5}$	\\ \hline
		$L_{c,\mathrm{LP}}$		&	$1.0\times 10^{-5}$	\\ \hline
		$s$			&	$2.0 \times 10^{-5}$	\\ \hline
		Mesh Size	&	\multirow{2}{*}{$1.0 \times 10^{-6}$} \\
		(Locally Refined) & \\ \hline
	\end{tabular}
	\end{center}
\end{table}

The choice of crack size given in Tables SM-2 and SM-3 is somewhat arbitrary and motivated by computational constraints, where it is necessary to choose a crack length several times larger than the characteristic size of the elements, and also significantly smaller than the characteristic length scales in the problem. The smallest of these length scales is the corner radius at the low-pressure end, which in general is not well-known. However, in these simulations, we choose it to be $r_{c} = 0.188$ mm, which is significantly smaller than the smallest characteristic gap size, $H - h = 2.5$ mm, yet still large enough to be significantly larger than the arbitrarily chosen characteristic crack length of 10 $\mu$m and characteristic element size of 1 $\mu$m.

\begin{figure}[h!t]
	\centerline{\includegraphics[scale=0.6]{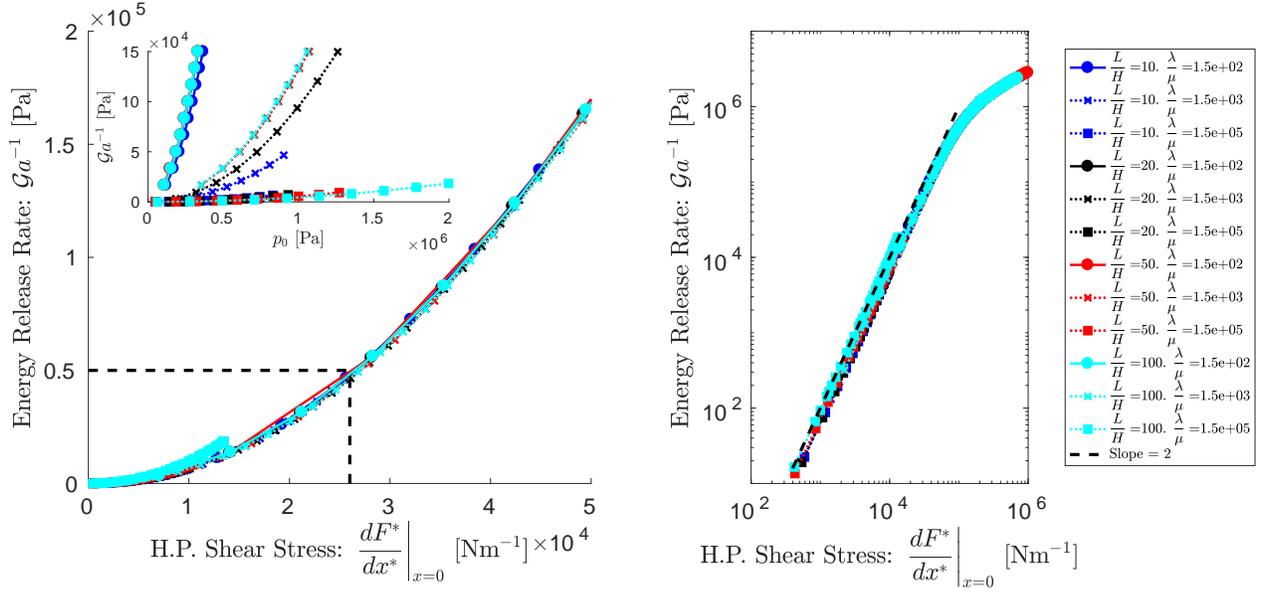}}
	\caption{Energy release rate versus shear stress at the leading edge. The shear stress is calculated entirely from the analytic model, with the lone fitting parameter being the low-pressure support stiffness, $k^{*}$, which was inferred independently. The point corresponding to the energy release rate of $\mathcal{G} = 0.5$ J/m$^{2}$ is denoted on the plot because it is used as the failure criterion in Fig. 7 of the paper. The right plot shows that the slope, for modest pressures, is 2 as predicted by linear elastic fracture mechanics.}
	\label{Fig: HP ERR - Gap Rat 090}
\end{figure}

As stated in the text, Fig. 7 is a plot of computational results for the smaller extrusion gap, where a failure criterion of $\mathcal{G}_{\mathrm{crit}} = 0.5$ J/m$^{2}$ for a crack length of $L_{c} = 10$ $\mu$m was chosen. The critical stress levels corresponding to this critical energy release rate are obtained from figures analogous to Figs. 5 and 6 in the paper (which were plots for the larger 25 percent extrusion gap). Fig. \ref{Fig: HP ERR - Gap Rat 090} is a plot of energy release rate versus analytically predicted axial stress gradient on the high-pressure end, and is analogous to Fig. 4 in the article. Fig. \ref{Fig: LP ERR - Gap Rat 090} in the Supplementary Information is a plot of energy release rate versus transmitted axial force on the low-pressure end, and is analogous to Fig. 5 in the article. Together, these two plots were used to determine the critical stresses corresponding to the critical energy release rates for the purposes of computing Fig. 7.
\begin{figure}[h!t]
	\centerline{\includegraphics[scale=0.6]{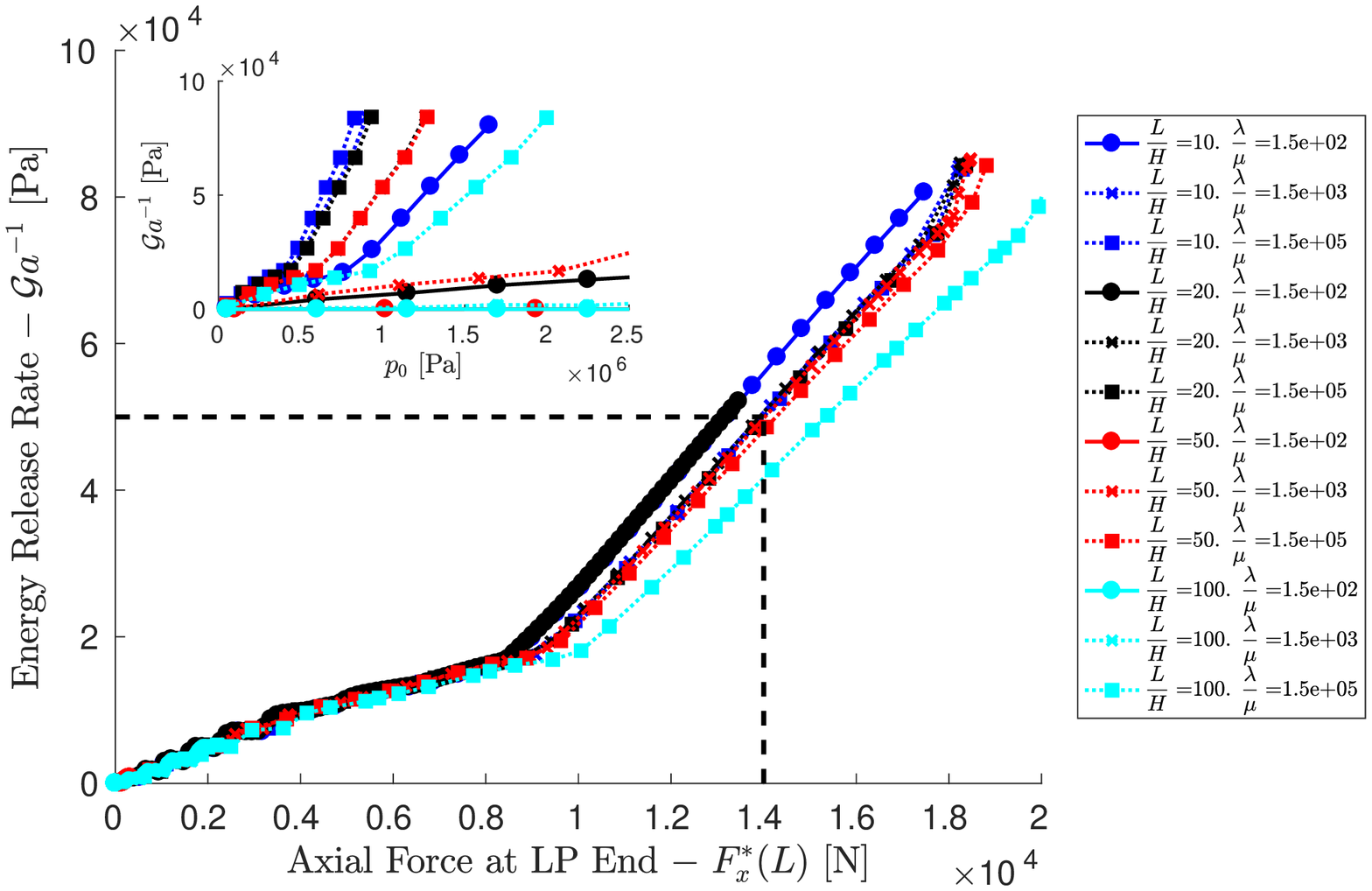}}
	\caption{Energy release rate versus axial force at the low-pressure end. The axial force is calculated entirely from the analytic model, with the lone fitting parameter being the low-pressure support stiffness, $k^{*}$, which was calculated independently. The point corresponding to the energy release rate of $\mathcal{G} = 0.5$ J/m$^{2}$ is denoted on the plot because it is used as the failure criterion in Fig. 7 of the paper.}
	\label{Fig: LP ERR - Gap Rat 090}
\end{figure}